\documentclass[twocolumn,superscriptaddress]{revtex4-2}
\usepackage{graphicx}
\usepackage{units}
\usepackage{multirow}
\usepackage{bigstrut}
\usepackage{overpic}
\usepackage{array}
\usepackage{color}
\usepackage{amsmath}
\usepackage{xspace}
\usepackage{amsfonts}
\usepackage{subfigure}
\usepackage{verbatim}
\usepackage{footmisc}
\usepackage{color}
\usepackage{epstopdf}
\usepackage[colorlinks,
            linkcolor=blue,
            anchorcolor=blue,
            citecolor=blue]{hyperref}
\RequirePackage{lineno}
\usepackage{appendix}

\newcommand{\dst}{D^{*}}
\newcommand{\dstbar}{\bar{D}^{*}}
\newcommand{\dstzero}{D^{*0}}
\newcommand{\dstplus}{D^{*+}}
\newcommand{\dstminus}{D^{*-}}

\newcommand{\dminus}{D^{-}}

\newcommand{\dzero}{D^{0}}
\newcommand{\ee}{e^+e^-}
\newcommand{\kaonp}{K^+}

\newcommand{\kp}{K^+}
\newcommand{\km}{K^-}

\newcommand{\piplus}{\pi^+}
\newcommand{\pip}{\pi^+}
\newcommand{\pim}{\pi^-}
\newcommand{\pizero}{\pi^0}

\newcommand{\jpsi}{J/\psi}
\newcommand{\ev}{\,\unit{eV}}
\newcommand{\mev}{\,\unit{MeV}}
\newcommand{\mevc}{\,\unit{MeV}/c}
\newcommand{\mevcc}{\,\unit{MeV}/c^2}
\newcommand{\gev}{\,\unit{GeV}}
\newcommand{\gevc}{\,\unit{GeV}/c}
\newcommand{\gevcc}{\,\unit{GeV}/c^2}
\newcommand{\invpb}{\,\unit{pb}^{-1}}
\newcommand{\invfb}{\,\unit{fb}^{-1}}

\begin{document}


\title{\boldmath Observation of Three Charmoniumlike States with $J^{PC}=1^{--}$ in $e^{+}e^{-}\to D^{*0}D^{*-}\pi^{+}$}

\date{\it \small \bf \today}

\author{
\begin{small}
\begin{center}
M.~Ablikim$^{1}$, M.~N.~Achasov$^{13,b}$, P.~Adlarson$^{73}$, R.~Aliberti$^{34}$, A.~Amoroso$^{72A,72C}$, M.~R.~An$^{38}$, Q.~An$^{69,56}$, Y.~Bai$^{55}$, O.~Bakina$^{35}$, I.~Balossino$^{29A}$
% ...
\\
\vspace{0.2cm}
(BESIII Collaboration)\\
\vspace{0.2cm} {\it
$^{1}$ Institute of High Energy Physics, Beijing 100049, People's Republic of China\\
% ...
$^{13}$ G.I. Budker Institute of Nuclear Physics SB RAS (BINP), Novosibirsk 630090, Russia\\
% ...
$^{29}$ INFN Sezione di Ferrara, (A)INFN Sezione di Ferrara, I-44122, Ferrara, Italy; (B)University of Ferrara,  I-44122, Ferrara, Italy\\
% ...
$^{34}$ Johannes Gutenberg University of Mainz, Johann-Joachim-Becher-Weg 45, D-55099 Mainz, Germany\\
$^{35}$ Joint Institute for Nuclear Research, 141980 Dubna, Moscow region, Russia\\
% ...
$^{38}$ Liaoning Normal University, Dalian 116029, People's Republic of China\\
% ...
$^{55}$ Southeast University, Nanjing 211100, People's Republic of China\\
$^{56}$ State Key Laboratory of Particle Detection and Electronics, Beijing 100049, Hefei 230026, People's Republic of China\\
% ...
$^{69}$ University of Science and Technology of China, Hefei 230026, People's Republic of China\\
% ...
$^{72}$ University of Turin and INFN, (A)University of Turin, I-10125, Turin, Italy; (B)University of Eastern Piedmont, I-15121, Alessandria, Italy; (C)INFN, I-10125, Turin, Italy\\
$^{73}$ Uppsala University, Box 516, SE-75120 Uppsala, Sweden\\
% ...
$^{b}$ Also at the Novosibirsk State University, Novosibirsk, 630090, Russia\\
% ...
}
\end{center}
\end{small}
}

\vspace{4cm}

\begin{abstract}
The Born cross sections of the process $e^{+}e^{-}\to D^{*0}D^{*-}\pi^{+}$ at center-of-mass energies from 4.189 to 4.951 GeV are measured for the first time. The data samples used correspond to an integrated luminosity of $17.9\,{\rm fb}^{-1}$ and were collected by the BESIII detector operating at the BEPCII storage ring. Three enhancements around 4.20, 4.47 and 4.67 GeV are visible. The resonances have masses of $4209.6\pm4.7\pm5.9\,{\rm MeV}/c^{2}$, $4469.1\pm26.2\pm3.6\,{\rm MeV}/c^{2}$ and $4675.3\pm29.5\pm3.5\,{\rm MeV}/c^{2}$ and widths of $81.6\pm17.8\pm9.0\,{\rm MeV}$, $246.3\pm36.7\pm9.4\,{\rm MeV}$, and $218.3\pm72.9\pm9.3\,{\rm MeV}$, respectively, where the first uncertainties are statistical and the second systematic. The first and third resonances are consistent with the $\psi(4230)$ and $\psi(4660)$ states, respectively, while the second one is compatible with the $\psi(4500)$ observed in the $e^{+}e^{-}\to K^{+}K^{-}J/\psi$ process. These three charmoniumlike $\psi$ states are observed in $e^{+}e^{-}\to D^{*0}D^{*-}\pi^{+}$ process for the first time.
\end{abstract}


\maketitle
The standard model of particle physics describes how quarks interact with each other to create various states of matter and antimatter.
Over the past years, a series of charmoniumlike vector meson $\psi$ states (also denoted as $Y$), containing at least $c\bar{c}$ quark pairs, have been observed via electron-positron annihilation in numerous experiments~\cite{Workman:2022ynf}.
Dedicated studies of these charmoniumlike states, which have unit spin and negative charge-parity quantum numbers $J^{PC}=1^{--}$, were initially triggered by the discovery of the $\psi(4230)$, previously called the $\psi(4260)$, in the $\ee\to\pip\pim\jpsi$ process at
\textit{BABAR}~\cite{ref:babar_4260:BaBar:2005hhc}, which was confirmed at
CLEO~\cite{ref:cleo_4260:CLEO:2006tct},
Belle~\cite{ref:belle_4260:Belle:2007dxy} and
BESIII~\cite{ref:bes_pipijpsi_1:BESIII:2016bnd,ref:bes_pipijpsi_2:Ablikim:2020pzw,ref:bes_pipijpsi_1:BESIII:2022jsj}.
Later, the $\psi(4360)$ and $\psi(4660)$ states were established in $\ee\to\pip\pim\psi(2S)$ at
\textit{BABAR}~\cite{ref:babar_pipipsi2s_1:BaBar:2006ait,ref:babar_pipipsi2s_2:BaBar:2012hpr},
Belle~\cite{ref:belle_pipipsi2s_1:Belle:2007umv,ref:belle_pipipsi2s_2:Belle:2014wyt}
and
BESIII~\cite{ref:bes_pipipsi2s_1:BESIII:2017tqk,ref:bes_pipipsi2s_2:BESIII:2017vtc,ref:bes_pipipsi2s_3:BESIII:2021njb}.
Some similar resonance enhancements around 4.23\,GeV, 4.36\,GeV or 4.66\,GeV are also reported in
$\pip\pim h_{c}$~\cite{ref:bes_pipihc:BESIII:2016adj},
$\omega\chi_{c0}$~\cite{ref:bes_omchicj:BESIII:2014rja,ref:bes_omchic0:BESIII:2019gjc},
$\eta\jpsi$~\cite{ref:bes_etajpsi:BESIII:2020bgb},
$\eta^{\prime}\jpsi$~\cite{ref:bes_etaprimejpsi:BESIII:2019nmu},
$\dzero\dstminus\pip$~\cite{ref:bes_pid0dst:BESIII:2018iea},
$\pi\pi\psi_2(3823)$~\cite{BESIII:2022yga},
$\pi^{+}\pi^{-}D^{+}D^{-}$~\cite{ref:bes_pipiDD:BESIII:2019phe,ref:bes_pipiDD:BESIII:2022quc}
and $\kp\km\jpsi$~\cite{ref:bes_kkjpsi:BESIII:2022joj} final states at BESIII. 
In addition, a new $\psi$ state, the $\psi(4500)$, was observed in $\ee\to K^+K^-\jpsi$, recently~\cite{ref:bes_kkjpsi:BESIII:2022joj}.
The genuine properties of these charmoniumlike $\psi$ states are still unknown, and there exist various theoretical interpretations, including tetraquarks, hybrid mesons, hadron molecules, hadrocharmonium, vector charmonia and threshold effects~\cite{ref:review_1:Chen:2016qju,ref:review_2:Guo:2017jvc}.

One striking feature of these charmoniumlike states is their large coupling to charmonium final states~\cite{Yuan:2021wpg}.
In contrast, there is a dip around the known $\psi(4230)$ mass in the cross section of $\ee\to\text{inclusive hadrons}$~\cite{ref:bes_xs_r-value} and in the
exclusive two-body production of $\ee\to D^{(*)}\bar{D}^{(*)}$~\cite{Briceno:2015rlt,BESIII:2021yvc}.
This is opposite to the behavior of conventional $c\bar{c}$ charmonium states lying above the $D\bar{D}$ mass threshold, which predominantly decays to open-charm final states~\cite{Workman:2022ynf}.
Therefore, it is essential to investigate the coupling of the charmoniumlike states to different open-charm channels to help identify their nature.

In the process $\ee\to\dzero\dstminus\pip$~\cite{ref:bes_pid0dst:BESIII:2018iea}, BESIII first determined a sizable coupling of the $\psi(4230)$ with the open-charm $\dzero\dstminus\pip$ decay, which is consistent with the hypothesis of a $D_1(2420)\bar{D}$ molecular state~\cite{ref:d2420d_1:Cleven:2013mka,ref:d2420d_2:Ding:2008gr,ref:d2420d_3:Li:2013yla,ref:d2420d_4:Li:2013bca,ref:d2420d_5:Wang:2013cya,ref:d2420d_6:Wu:2013onz,ref:d2420d_7:Ji:2022blw}.
In lattice QCD, the leptonic partial width $\Gamma^{ee}_{\psi(4230)}$ of the $\psi(4230)$ is predicted to be less than $40\ev$ using a hybrid scenario~\cite{ref:lqcd_2016:Chen:2016ejo}.
To date, $\Gamma^{ee}_{\psi(4230)}$ is evaluated to be $36.4\pm4.7\ev$, based on the combined analysis of the known $\psi(4230)$ decay channels~\cite{ref:Gee4260_2018:Zhang:2018zog}.
Study of the $\psi(4230)$ in the open-charm process $\ee\to\dst\dstbar\pi$ provides new input to $\Gamma^{ee}_{\psi(4230)}$ and to the relative size of different decay modes, which can be used to test the theoretical explanation of the hybrid and $D \bar{D}_1$ molecular state models.

The $\psi(4500)$ is reported in the $\ee\to K^+K^-\jpsi$ process~\cite{ref:bes_kkjpsi:BESIII:2022joj}.
Its spin-parity and mass agree with 
the lattice QCD calculation for a $c\bar{c}s\bar{s}$ tetraquark state~\cite{Chiu:2005ey};
a baryonium state~\cite{Qiao:2007ce};
a $D^{*}\bar{D}_{2}$ molecule state~\cite{Dong:2021juy};
a hidden-charm tetraquark candidate in QCD sum rule~\cite{Wang:2021qus};
a $D_{s}\bar{D}_{s1}$ molecule state~\cite{Dong:2021juy,
Peng:2022nrj}, which is a hidden-strangeness partner of the $\psi(4230)$ state under the $D \bar{D}_1$ molecule assumption~\cite{Peng:2022nrj}, and a vector charmonium $5S$-$4D$ mixing state~\cite{Wang:2022jxj}.
To explore its true nature, an independent confirmation of the $\psi(4500)$ in another channel would be important and crucial.
In addition, since the Born cross section of $\ee\to K Z_{cs}(3985)$ peaks around the $\psi(4660)$ mass~\cite{BESIII:2020qkh,BESIII:2022qzr}, the $\psi(4660)$ is considered to be a hidden-strangeness state. 
Given that the heavier $\psi(4500)$ and $\psi(4660)$ states have not been observed in open-charm decay, it is also highly desirable to explore these states in the process $\ee\to\dst\dstbar\pi$ to help determine their quark constituents.

In this Letter, the Born cross sections of the $\ee\to\dstzero\dstminus\pip$ processes are measured at 86 center-of-mass energies from 4.189 to 4.951\,GeV for the first time.
Charge conjugate modes are always implied throughout this Letter.
The datasets used are accumulated with the BESIII detector at the BEPCII collider and correspond to an integrated luminosity of $17.9\invfb$~\cite{ref:lumi_eng_scan:BESIII:2017lkp,ref:lumi_eng_xyz_1:BESIII:2020eyu,ref:lumi_eng_xyz_2:BESIII:2022ulv}.
Details about BEPCII and BESIII can be found in Refs.~\cite{Yu:IPAC2016-TUYA01, Ablikim:2009aa, Ablikim:2019hff}.
The datasets include 49 energy points with integrated luminosities less than $10\invpb$ (``scan data'') and another 37 energy points with larger integrated luminosities (``\textit{XYZ} data'').
Details of the datasets can be found in Tables~I and II of the Supplemental Material~\cite{Supplemental}.
By fitting the line shape of dressed cross sections, which takes into account vacuum polarizations~\cite{Jin:2018kjv}, we report the observation of three vector charmoniumlike $\psi$ states in $\ee\to\dstzero\dstminus\pip$.

Simulated data samples are produced with {\sc geant4}-based~\cite{geant4} Monte Carlo (MC) software, which includes the geometric description~\cite{Huang:2022wuo} of the BESIII detector and the detector response, as detailed in Ref.~\cite{Ablikim:2019hff}.
The simulation models the beam energy spread and initial state radiation~(ISR) in the $\ee$ annihilation with the generator {\sc kkmc}~\cite{kkmc}.
The signal MC samples of the $\ee\to\dstzero\dstminus\pip$ process are generated according to the partial-wave-analysis results at each energy point.
Possible background contributions are estimated by inclusive MC simulation samples, which include the production of open-charm processes, the ISR production of vector charmonium(like) states, and the continuum processes incorporated in {\sc kkmc}.
All particle decays are modeled with {\sc evtgen}~\cite{evtgen} using branching fractions~(BFs) taken from the Particle Data Group~(PDG)~\cite{Workman:2022ynf}, when available, and unknown $J/\psi$ and $\psi(2S)$ decays are estimated with {\sc lundcharm}~\cite{lundcharm}.
Final state radiation from charged final state particles is incorporated using {\sc photos}~\cite{photos}.

\begin{figure}[htp]
    \begin{center}
        \includegraphics[width=0.98\linewidth]{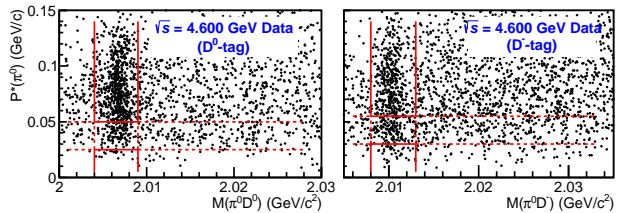}
        \caption{
            Two-dimensional distributions of $P^{*}(\pizero)$ versus $M(\pizero D)$ for the $\dzero$tag (left) and $\dminus$tag (right) methods in data at $\sqrt{s}=4.600\gev$.
            The events between the vertical solid lines are kept and those between the horizontal dashed lines are vetoed based on the requirements for $M(\pizero D)$ and $P^{*}(\pizero)$.
            The distributions for signal MC simulation samples are shown in Fig.~1 of the Supplemental Material~\cite{Supplemental}. }
		\label{fig:2d_p_mdst}
	\end{center}
\end{figure}

To improve the signal selection efficiency, a partial-reconstruction technique is employed to identify the $\dstzero\dstminus\pip$ final states, in which two tagging methods, the $\dzero$tag and the $\dminus$tag, are performed.
In the $\dzero(\dminus)$tag method, the bachelor charged $\pip$ from primary production, the $\dzero(\dminus)$ meson, and at least one soft $\pizero(\to\gamma\gamma)$ from $\dstzero(\dstminus)\to\dzero(\dminus)\pizero$ decay are reconstructed.
To improve the signal purity, only the decays $\dzero\to\km\pip$, $\km\pip\pizero$, and $\km\pip\pip\pim$ ($\dminus\to\kp\pim\pim$), which have relatively large BFs, are reconstructed.
By reconstructing the $\dstzero(\dstminus)$ and the bachelor $\pip$, the flavor of the missing $\dstminus(\dstzero)$ meson is fixed.
All the charged tracks and $\pizero$ candidates are selected following the criteria in Ref.~\cite{BESIII:2013mhi}.
To form candidates for $\dzero\to\km\pip$, $\dzero\to\km\pip\pizero$, $\dzero\to\km\pip\pip\pim$ and $\dminus\to\kaonp\pim\pim$ decays, the reconstructed final state invariant masses are required to be within $(1.835,1.887)$, $(1.827,1.882)$, $(1.855,1.874)$, and $(1.856,1.883)\gevcc$, respectively.
Here, the different mass regions are due to the various momentum resolutions.
The $\pizero$ candidates from $\dst$ decays can be either from the reconstructed or missing $\dst$ candidates.
For the $\pizero$ from missing $\dst$ candidates, its momentum in the reconstructed $D\pip$ recoil system, $P^{*}(\pizero)$, peaks around $40\mevc$.
To distinguish the source of $\pizero$ with reconstructed $\dst$ candidates, the reconstructed invariant masses are required to satisfy $M(\dzero\pizero)\in(2.004, 2.009)\gevcc$ with $P^{*}(\pizero)\notin(0.025, 0.050)\gevc$ in the $\dzero$tag method, and $M(\dminus\pizero)\in(2.008, 2.013)\gevcc$ with $P^{*}(\pizero)\notin(0.030, 0.055)\gevc$ in the $\dminus$tag method, as shown in Fig.~\ref{fig:2d_p_mdst} for data at $\sqrt{s}=4.600\gev$.
Moreover, the $\pip\dzero$ invariant mass must be greater than $2.02\gevcc$ in the $\dzero$tag method to reject background for the bachelor $\pip$ from $\dstplus\to\pip\dzero$.

To improve the resolution and further suppress the background, a kinematic fit (3C) is performed to constrain the reconstructed $\pizero$, $\dzero(\dminus)$, and $\dstzero(\dstminus)$ mesons to their individual known masses~\cite{Workman:2022ynf}.
Candidate events are required to have $\chi_{\rm 3C}^{2}<50$ and the fitted four-momenta of all related particles are used for further analysis.
If there is more than one $\pizero\dzero(\dminus)$ candidate in an event, only the one with the minimum $\chi_{\rm 3C}^{2}$ is retained.
Furthermore, if one event survives in both tag methods, only the combination in the $\dzero$tag method is kept to avoid double counting in the simultaneous fit.

\begin{figure}[tp]
	\begin{center}
		\includegraphics[width=0.98\linewidth]{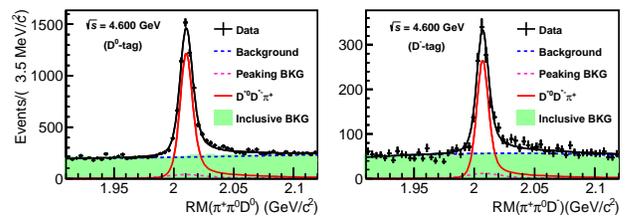}
		\caption{
            The distributions of the recoil masses $RM(\pip\pizero\dzero)$ (left) and $RM(\pip\pizero\dminus)$ (right) for data at $\sqrt{s}=4.600\gev$ with simultaneous fit results overlaid.
            The red solid-curve is the signal shape.
            The pink and blue dashed-curve are the peaking and smooth background, respectively.
            The light green shadowed histogram is the simulated inclusive background MC samples.
        }
		\label{fig:simufit_4600}
	\end{center}
\end{figure}

Figure~\ref{fig:simufit_4600} shows the distributions of the recoil masses of reconstructed $\pip$, $\pizero$ and $D$ mesons, $RM(\pip\pizero\dzero)$ and $RM(\pip\pizero\dminus)$.
The background study based on inclusive MC simulation samples shows that the shape of the background at each energy point is smooth and can be well described by a second-order Chebyshev function.
A peaking background is found in the signal MC simulation due to the miscombination of particles from the missing and tagged sides.
Its shape is obtained by selecting the unmatched events from inclusive MC simulation samples, in which the missing $\dst$ candidate decays inclusively while the tagged $\dst$ candidate decays into the signal process final state.
Its contribution is fixed in the fit according to the ratio between matched and unmatched events in the MC simulation.

An unbinned extended maximum likelihood fit is performed on the distributions of $RM(\pip\pizero\dzero)$ and $RM(\pip\pizero\dminus)$ simultaneously to determine the Born cross section at each energy point.
Figure~\ref{fig:simufit_4600} shows the fit results at $\sqrt{s}=4.600\gev$, as an example.
The signal shape is derived from MC simulation convolved with a Gaussian function with free parameters to account for the resolution difference between data and MC simulation.
The background shape is parametrized as a sum of the shape from unmatched MC samples and a second-order Chebyshev function.
The Born cross sections ($\sigma^{\rm Born}$) at the individual energy points are defined as
\begin{eqnarray}
    \begin{aligned}
        \sigma^{\rm Born}=&\frac{\sigma^{\rm dressed}}{\frac{1}{|1-\Pi|^{2}}}\\=&\frac{N^{\rm obs}_{D^{0(-)}\text{tag}}}{\mathcal{L}_{\rm int}\cdot\epsilon_{D^{0(-)}\text{tag}}\cdot\hat{\mathcal{B}}_{D^{0(-)}\text{tag}}\cdot(1+\delta^{\rm ISR})\cdot\frac{1}{|1-\Pi|^{2}}}.\nonumber
    \end{aligned}
\end{eqnarray}
Here, $N^{\rm obs}_{D^{0(-)}\text{tag}}$ is calculated according to $\sigma^{\rm Born}$ which is taken as a common parameter in the simultaneous fit,
$\epsilon_{{D^{0(-)}\text{tag}}}$ is the detection efficiency,
$\mathcal{L}_{\rm int}$ is the integral luminosity measured by Refs.~\cite{ref:lumi_eng_scan:BESIII:2017lkp,ref:lumi_eng_xyz_1:BESIII:2020eyu,ref:lumi_eng_xyz_2:BESIII:2022ulv},
$\hat{\mathcal{B}}_{{D^{0(-)}\text{tag}}}$ stands for an equivalent BF including all the related products of the BF obtained from the PDG~\cite{Workman:2022ynf}, while $(1+\delta^{\rm ISR})$ and
$(1/|1-\Pi|^{2})$ are the correction factors for ISR and vacuum polarization~\cite{Jegerlehner:2011mw}.
To estimate the ISR factors and consider the correlation effect on detection efficiencies, an iterative weighting method~\cite{Sun:2020ehv} is performed to correct the corresponding dressed cross section values.
All the numerical results from the fits are summarized in Tables~I and II of the Supplemental Material~\cite{Supplemental} for the \textit{XYZ} and scan data samples, respectively.

The systematic uncertainties in the Born cross section measurements, as detailed in Supplemental Material~\cite{Supplemental}, are divided into three parts.
The first part relates to the determination of the detection efficiency, including the tracking, particle identification, $\pizero$ reconstruction, signal region requirements, signal decay model and ISR correction factor.
The second part relates to the estimation of signal yields from the fit, consisting of the signal and background shapes as well as the fit range.
The last part includes the uncertainties from the luminosities and the intermediate BFs.
The items in the first and third parts are completely correlated between different energy points, except for the uncertainties due to signal region requirements and the signal decay model.
For the second part at low-yield ($<300$ events) energy points, the systematic uncertainties obtained at their nearest energy point in high-yield ($>300$ events) \textit{XYZ} data are used.
All the systematic uncertainties are studied for each tag method and combined to obtain the total systematic uncertainties according to their signal yields. 
The total relative systematic uncertainties at different energy points are between $6.7$ and $9.6\%$.

\begin{figure}[h]
	\begin{center}
        \includegraphics[width=0.98\linewidth]{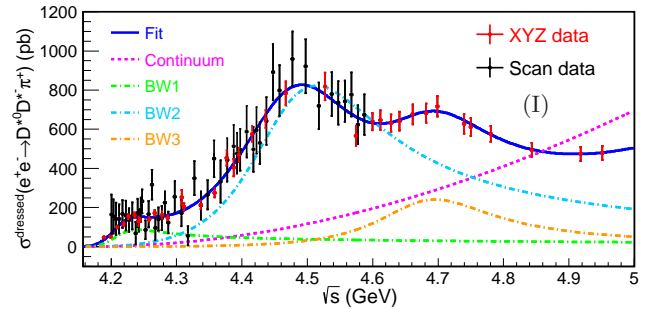}\put(-45,75){(I)}
        \caption{
            The fit results (solution I) of the dressed cross section line shape of $\ee\to\dstzero\dstminus\pip$.
            The black and red points with error bars are data, including statistical and systematic uncertainties.
            The blue curve is the total fit.
            The green, azure and orange dashed curves describe three BW functions, and the pink dashed curve is the three body phase space contribution.}
		\label{fig:LS_fit}
	\end{center}
\end{figure}

The dressed cross sections obtained at various energy points are shown in Fig.~\ref{fig:LS_fit}.
Three possible enhancements around 4.20, 4.47, and 4.67\,GeV are observed.
To fit this line shape, we use the coherent sum of a continuum amplitude for $\ee\to\dstzero\dstminus\piplus$ and three resonance amplitudes described by relativistic Breit-Wigner~(BW) functions
\begin{eqnarray}
    \sigma^{\rm dressed}(\sqrt{s})=C_{0}\Big|C_{1}\sqrt{\Phi(\sqrt{s})}+\sum^{3}_{k=1}{\rm BW}_{k}(\sqrt{s})e^{i\phi_{k}}\Big|^2,\nonumber
    \label{eq:lineshape}
\end{eqnarray}
where $C_{0}=3.894\times10^{5}\,\unit{nb}\cdot\gev^{2}$ is a unit conversion factor, 
$C_{1}$ is the continuum free parameter, 
and $\phi_{k}$ is the phase angle among different components.
The relativistic BW amplitude for a resonance $R_{k}\to\dstzero\dstminus\pip$ is written as
\begin{eqnarray}
    {\rm BW}_{k}(\sqrt{s})=\frac{m_{k}}{\sqrt{s}}\cdot\frac{\sqrt{12\pi\cdot\Gamma^{ee}_{k}\cdot\mathcal{B}_k\cdot\Gamma^{\rm tot}_{k}}}{s-m_{k}^{2}+im_{k}\Gamma^{\rm tot}_{k}}\cdot\sqrt{\frac{\Phi(\sqrt{s})}{\Phi(m_{k})}},\nonumber
    \label{eq:bw}
\end{eqnarray}
where $m_{k}$ and $\Gamma^{\rm tot}_{k}$ are the $k$th resonance mass and total width, respectively,
$\Gamma^{ee}_{k}\cdot\mathcal{B}_k$ is the leptonic width of the $k$th resonance times the BF of
$R_{k}\to\dstzero\dstminus\pip$,
and $\Phi(\sqrt{s})$ is the three body phase space contribution defined as
$\Phi(\sqrt{s})=\iint[1/(2\pi)^{3}32(\sqrt{s})^3]dm_{23}^{2}dm_{12}^{2}$~\cite{Workman:2022ynf}.

The $\chi^{2}$ of the fit to the dressed cross section line shape is constructed according to the method in Ref.~\cite{BESIII:2020kpr} by incorporating both the statistical and systematic uncertainty and considering both the correlated and uncorrelated terms.
To avoid biasing the $\chi^{2}$ minimization, the correlated uncertainties are calculated according to the predicted cross section values times the corresponding relative uncertainties when constructing the covariance matrix~\cite{Sun:2005ip}.

The fit result is shown in Fig.~\ref{fig:LS_fit}.
There are eight solutions with the same fit quality with identical continuum contributions as well as masses and widths for the resonances~\cite{Bai:2019jrb}.
However, the resulting product $\Gamma^{ee}_{k}\mathcal{B}_k$ and phases $\phi_{k}$ are different, as plotted in Fig.~2 of Supplemental Material~\cite{Supplemental}.
The numerical results are listed in Table~\ref{tab:LS_parameters}.
In general, the magnitudes of $\Gamma^{ee}_{k}\mathcal{B}_k$ become increased when the destructive interference effects due to relative phase angles are larger.
The dressed cross sections are also fitted under the assumption of only two resonances plus the continuum component.
The relative changes in the $\chi^{2}$ value ($\Delta{\rm \chi^{2}}=130.5$) and the number of degrees of freedom ($\Delta{\rm ndof}=4$) are used to estimate the significance of the three-resonance hypothesis over the two-resonance hypothesis as $10.8\sigma$.
The significance of the two-resonance hypothesis over the one-resonance hypothesis is $22.8\sigma$ according to the changes of $\Delta{\rm \chi^{2}}=537.1$ and $\Delta{\rm ndof}=4$.

\begin{table*}[tbp]
    \begin{center}
        \caption{
            The fit results of the dressed cross section line shape of $\ee\to\dstzero\dstminus\pip$ with eight different solutions, which have the same fit quality as shown in Fig.2 in Supplemental Material~\cite{Supplemental}.
            For the uncertainties of the masses and widths, those from the solutions with maximum uncertainties are adopted.}
        \begin{tabular}{l c c c c c c c c}
        \hline \hline
        &I &II &III &IV &V &VI &VII &VIII \\ \hline
        $C_{1}\,(10^{-3})$                   &\multicolumn{8}{c}{$4.2\pm1.5$}    \\ \hline
        $m_{1}\,(\unit{MeV}/c^2)$            &\multicolumn{8}{c}{$4209.6\pm4.7$} \\
        $\Gamma^{\rm tot}_{1}\,(\unit{MeV})$ &\multicolumn{8}{c}{$81.6\pm17.8$} \\
        $\Gamma^{ee}_{1}\mathcal{B}_1\,(\unit{eV})$ 
            &$5.4\pm1.1$ &$6.0\pm1.3$ &$4.8\pm0.9$ &$5.3\pm1.1$ &$17.9\pm7.2$ &$19.8\pm6.6$ &$20.2\pm7.4$ &$22.4\pm9.0$ \\
        $\phi_{1}\,(\unit{rad})$
            &$3.1\pm0.5$ &$3.8\pm0.4$ &$1.9\pm0.7$ &$2.6\pm0.6$ &$4.2\pm0.3$  &$4.8\pm0.2$  &$5.4\pm0.3$  &$6.0\pm0.3$  \\ 
        \hline
        $m_{2}\,(\unit{MeV}/c^2)$            &\multicolumn{8}{c}{$4469.1\pm26.2$} \\
        $\Gamma^{\rm tot}_{2}\,(\unit{MeV})$ &\multicolumn{8}{c}{$246.3\pm36.7$} \\
		$\Gamma^{ee}_{2}\mathcal{B}_2\,(\unit{eV})$
            &$243.3\pm83.5$ &$832.5\pm716.5$ &$107.4\pm50.6$ &$367.4\pm370.8$ &$225.5\pm94.9$ &$770.8\pm383.8$ &$510.1\pm202.3$ &$1744.3\pm926.9$ \\
		$\phi_{2}\,(\unit{rad})$
            &$4.4\pm0.3$    &$-0.9\pm0.3$    &$2.6\pm0.6$    &$3.7\pm0.8$     &$1.9\pm0.8$    &$3.0\pm0.4$     &$3.7\pm0.3$     &$-1.5\pm0.3$     \\
		\hline
		$m_{3}\,(\unit{MeV}/c^2)$            &\multicolumn{8}{c}{$4675.3\pm29.5$} \\
		$\Gamma^{\rm tot}_{3}\,(\unit{MeV})$ &\multicolumn{8}{c}{$218.3\pm72.9$} \\
		$\Gamma^{ee}_{3}\mathcal{B}_3\,(\unit{eV})$
            &$75.8\pm148.8$ &$1601.9\pm1152.6$ &$19.4\pm27.1$ &$411.6\pm230.5$ &$24.4\pm34.5$ &$515.6\pm244.6$ &$95.1\pm173.1$ &$2005.3\pm1166.1$ \\
		$\phi_{3}\,(\unit{rad})$                    
            &$4.9\pm1.4$    &$-2.9\pm0.4$      &$2.1\pm0.4$   &$0.6\pm1.1$     &$1.7\pm0.5$   &$6.5\pm0.5$     &$4.5\pm1.3$    &$-3.3\pm0.3$      \\
		\hline\hline
        \end{tabular}
        \label{tab:LS_parameters}
    \end{center}
\end{table*}

The systematic uncertainties of the resonance parameters are dominated by those from the center-of-mass energy calibration, beam energy spread, and parametrization of the continuum contribution.
Other uncertainties from the measured cross sections have been included in the line shape fit.
The uncertainty from the center-of-mass energy measurement is estimated by propagating the largest uncertainty of the measured energies ($0.8\mevcc$) to the $\psi$-state mass parameter.
The uncertainty from beam energy spread is considered by smearing the energy with its spread value at each energy point.
The differences of resonance parameters determined from fits using nominal and smeared line shapes are taken as the systematic uncertainties.
To estimate the uncertainty related to the fit model, the three body continuum contribution is replaced by a third-order polynomial parametrized function.
The resulting differences in the masses and widths of resonances are taken as systematic uncertainties.
The total systematic uncertainty is obtained by summing the individual values in quadrature, assuming they are all uncorrelated, as listed in Table~\ref{tab:res_sys_un}.

\begin{table}[tbp]
	\begin{center}
		\caption{
            The systematic uncertainties in the measurements of the $\psi$-state parameters.
            }
            \begin{tabular}{l c c c|c}
            \hline\hline
			Source                       &Energy   &Beam spread &Fit model &Total \\\hline
            $m_{1}$$(\mevcc)$            &0.8      &5.5         &2.0       &5.9   \\
            $\Gamma_{1}^{\rm tot}(\mev)$ &$\cdots$ &1.7         &8.8       &9.0   \\\hline
            $m_{2}$$(\mevcc)$            &0.8      &3.5         &0.7       &3.6   \\
            $\Gamma_{2}^{\rm tot}(\mev)$ &$\cdots$ &6.9         &6.4       &9.4   \\\hline
            $m_{3}$$(\mevcc)$            &0.8      &1.5         &3.1       &3.5   \\
            $\Gamma_{3}^{\rm tot}(\mev)$ &$\cdots$ &7.4         &5.7       &9.3   \\\hline
			\hline
			\end{tabular}
		\label{tab:res_sys_un}
	\end{center}
\end{table}

In summary, the Born cross sections of the process $\ee\to\dstzero\dstminus\piplus$ at 86 center-of-mass energies from $\sqrt{s}=4.189$ to $4.951\gev$ are measured for the first time with the data samples collected by the BESIII detector.
Fitting the dressed cross sections with a three-resonance hypothesis, their masses and widths are determined to be
$m_{1}=4209.6\pm4.7\pm5.9\mevcc$, and $\Gamma_{1}=81.6\pm17.8\pm9.0\mev$ [denoted as $\psi(4210)$],
$m_{2}=4469.1\pm26.2\pm3.6\mevcc$, and $\Gamma_{2}=246.3\pm36.7\pm9.4\mev$ [denoted as $\psi(4470)$],
$m_{3}=4675.3\pm29.5\pm3.5\mevcc$, and $\Gamma_{3}=218.3\pm72.9\pm9.3\mev$ [denoted as $\psi(4660)$], 
where the first uncertainties are statistical and the second are systematic.
The significance of the three-resonance hypothesis compared with the two-resonance one is greater than $10\sigma$.
The mass of $\psi(4210)$ is consistent with the mass of $\psi(4230)$ from the combined fit in Ref.~\cite{ref:Gee4260_2018:Zhang:2018zog}.
If we assume they are the same resonance, $\Gamma^{ee}_{\psi(4230)}$ becomes greater than 40 eV, which disfavors the hybrid interpretation under the lattice QCD calculation~\cite{ref:lqcd_2016:Chen:2016ejo}.
In addition, we find the couplings of $\psi(4230)$ to $\dstzero\dstminus\piplus$ and $\dzero\dstminus\pip$ are at the same order of magnitude.
This is the first observation of the state $\psi(4470)$ in an open-charm process, and its resonance parameters are compatible with those of the $\psi(4500)$ state observed in $\ee\to\kp\km\jpsi$~\cite{ref:bes_kkjpsi:BESIII:2022joj}.
Assuming the $\psi(4470)$ and $\psi(4500)$ are the same state, the rate of its decay to $\dst\dstbar\pi$ is 2 orders of magnitude greater than that to $K\bar{K}\jpsi$, which is inconsistent with the conjectured hidden-strangeness tetraquark nature of the $\psi(4500)$~\cite{Chiu:2005ey,Dong:2021juy, Peng:2022nrj}.
We confirm for the first time the existence of the resonance $\psi(4660)$ in open-charm final states with resonance parameters consistent with the latest results derived in $\ee\to\pi^+\pi^- \psi(2S)$ at BESIII~\cite{ref:bes_pipipsi2s_3:BESIII:2021njb}.
However, the relative size of their couplings cannot be constrained by current data, as different fit solutions result in large variations of the product $\Gamma^{ee}_{\psi(4660)}\mathcal{B}_{\psi(4660)}$.
Further amplitude analyses of different open- and hidden-charm final states are desired to advance our knowledge of the nature of these charmoniumlike $\psi$ states.

The BESIII collaboration thanks the staff of BEPCII and the IHEP computing center for their strong support. This work is supported in part by National Key R\&D Program of China under Contracts Nos. 2020YFA0406400, 2020YFA0406300; National Natural Science Foundation of China (NSFC) under Contracts Nos. 11635010, 11735014, 11835012, 11935015, 11935016, 11935018, 11961141012, 12022510, 12025502, 12035009, 12035013, 12061131003, 12192260, 12192261, 12192262, 12192263, 12192264, 12192265, 12221005; the Chinese Academy of Sciences (CAS) Large-Scale Scientific Facility Program; the CAS Center for Excellence in Particle Physics (CCEPP); Joint Large-Scale Scientific Facility Funds of the NSFC and CAS under Contract No. U1832207; CAS Key Research Program of Frontier Sciences under Contracts Nos. QYZDJ-SSW-SLH003, QYZDJ-SSW-SLH040; 100 Talents Program of CAS; Fundamental Research Funds for the Central Universities, Lanzhou University, University of Chinese Academy of Sciences; The Institute of Nuclear and Particle Physics (INPAC) and Shanghai Key Laboratory for Particle Physics and Cosmology; ERC under Contract No. 758462; European Union's Horizon 2020 research and innovation programme under Marie Sklodowska-Curie grant agreement under Contract No. 894790; German Research Foundation DFG under Contracts Nos. 443159800, 455635585, Collaborative Research Center CRC 1044, FOR5327, GRK 2149; Istituto Nazionale di Fisica Nucleare, Italy; Ministry of Development of Turkey under Contract No. DPT2006K-120470; National Research Foundation of Korea under Contract No. NRF-2022R1A2C1092335; National Science and Technology fund; National Science Research and Innovation Fund (NSRF) via the Program Management Unit for Human Resources \& Institutional Development, Research and Innovation under Contract No. B16F640076; Polish National Science Centre under Contract No. 2019/35/O/ST2/02907; Suranaree University of Technology (SUT), Thailand Science Research and Innovation (TSRI), and National Science Research and Innovation Fund (NSRF) under Contract No. 160355; The Royal Society, UK under Contract No. DH160214; The Swedish Research Council; U. S. Department of Energy under Contract No. DE-FG02-05ER41374.



\onecolumngrid

\newpage
\appendix
\setcounter{table}{0}
\setcounter{figure}{0}

{\large \bf Supplemental Material for ``Observation of Three Charmoniumlike States with $J^{PC}=1^{--}$ in $e^{+}e^{-}\to D^{*0}D^{*-}\pi^{+}$''}
\begin{appendices}
    \section{Details on event selection}
\begin{figure}[htp]
        \begin{center}
            \includegraphics[width=0.48\linewidth]{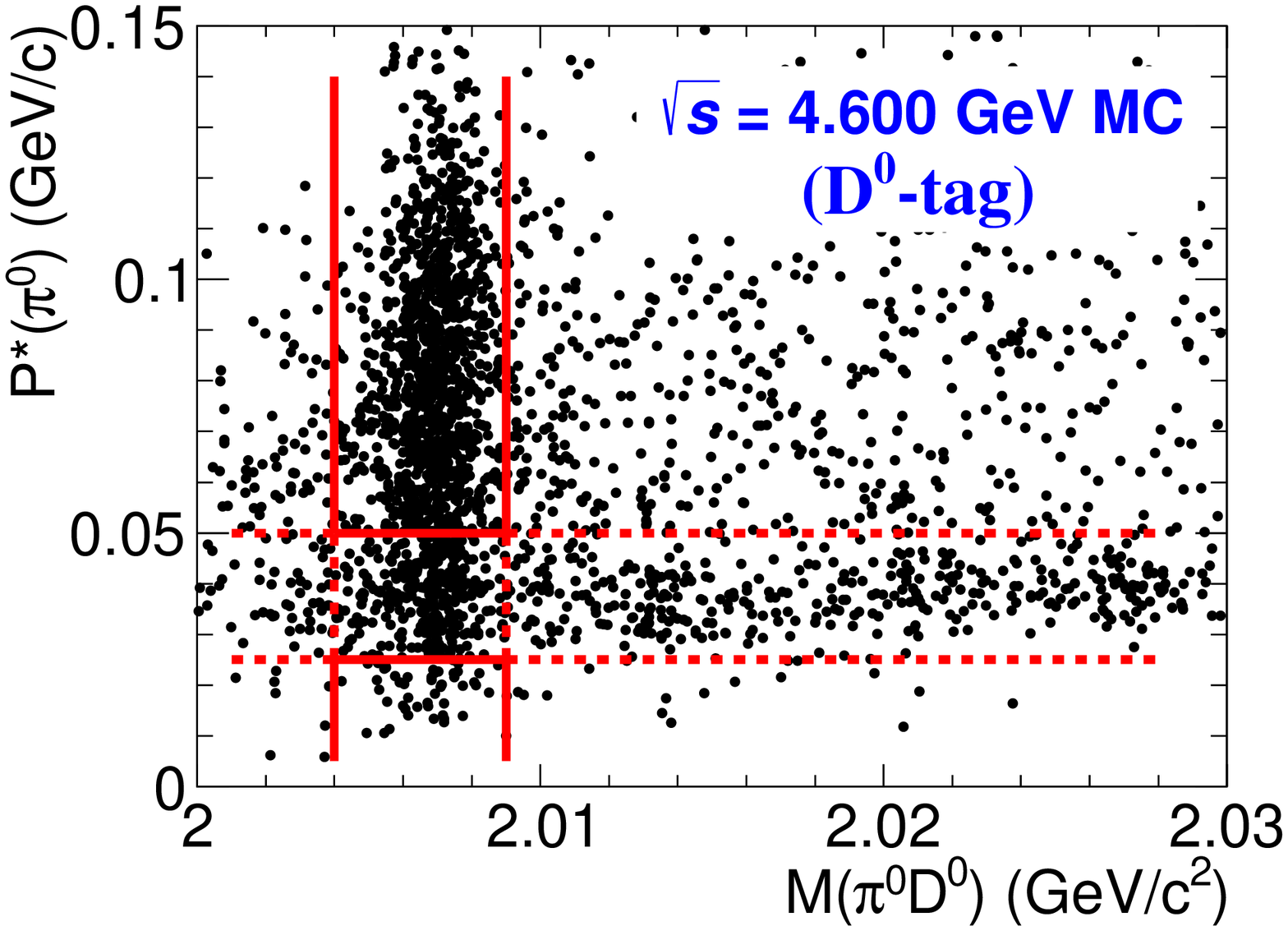}
            \includegraphics[width=0.48\linewidth]{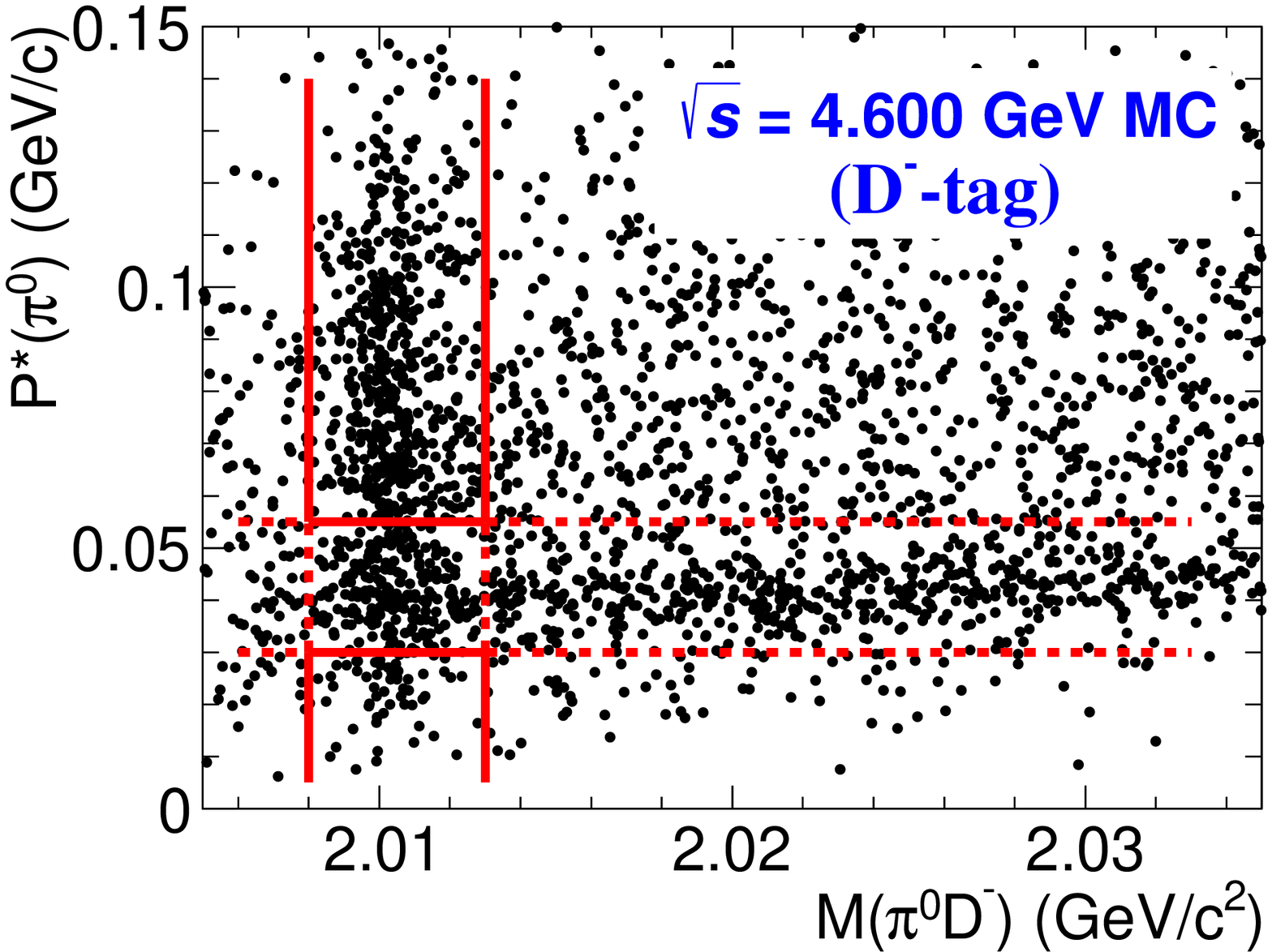}
            \caption{
                Two-dimensions distributions of $P^{*}(\pizero)$ versus $M(\pizero D)$ for MC simulation with $\dzero$tag and $\dminus$tag methods at $\sqrt{s}=4.600\gev$.
                The events are kept inside the vertical solid lines and vetoed inside the horizontal dashed lines based on the requirements for $M(\pizero D)$ and $P^{*}(\pizero)$.
            }
            \label{supp:fig:2d_p_mdst}
        \end{center}
    \end{figure}

    Two-dimensional distributions of $P^{*}(\pizero)$ versus $M(\pizero D)$ for MC samples with $\dzero$tag and $\dminus$tag methods at $\sqrt{s}=4.600\gev$ are shown in Fig.~\ref{supp:fig:2d_p_mdst}, respectively.
    
    \section{Signal yields and Born cross section}
    
    A simultaneous fit of $\dzero$tag and $\dminus$tag is performed at each energy point according to the calculation of the Born cross section.
    \begin{eqnarray}
    \begin{aligned}
    \sigma^{\rm Born}=\frac{\sigma^{\rm dressed}}{\frac{1}{|1-\Pi|^{2}}}=\frac{N^{\rm obs}_{D^{0(-)}\text{tag}}}{\mathcal{L}_{\rm int}\epsilon_{D^{0(-)}\text{tag}}\hat{\mathcal{B}}_{D^{0(-)}\text{tag}}(1+\delta^{\rm ISR})\frac{1}{|1-\Pi|^{2}}}.\nonumber
    \end{aligned}
    \end{eqnarray}
    The integral luminosities $\mathcal{L}_{\rm int}$ are measured by Refs.~\cite{ref:lumi_eng_scan:BESIII:2017lkp,ref:lumi_eng_xyz_1:BESIII:2020eyu,ref:lumi_eng_xyz_2:BESIII:2022ulv}.
    The $\sigma^{\rm Born}$ is taken as a common parameter in the fitting while detection efficiencies, ISR and vacuum polarization factors are estimated based on MC simulation.
    With the fit results shown in Tables~\ref{supp:tab:xs_xyz} and~\ref{supp:tab:xs_scan}, $N_{D^{0}\text{tag}}^{\rm obs}$ and $N_{D^{-}\text{tag}}^{\rm obs}$ can be calculated directly.
    
    \begin{table}[htp]
        \begin{center}
            \caption{
                The Born cross section of $\ee\to\dstzero\dstminus\pip$ for XYZ data sets, where the charge conjugation mode is also included.
                The first uncertainties are statistical and the second ones systematic.
                }
            \begin{tabular}{c c c c c c c c c c}
            \hline\hline
            &$\sqrt{s}$\,(GeV) &$\mathcal{L}_{\rm int}$\,(pb$^{-1}$) &$1+\delta^{\rm ISR}$ &$\frac{1}{|1-\Pi|^{2}}$ &$N_{D^{0}\text{tag}}^{\rm obs}$ &$\epsilon_{D^{0}\text{tag}}$\,(\%) &$N_{D^{-}\text{tag}}^{\rm obs}$ &$\epsilon_{D^{-}\text{tag}}$\,(\%) &$\sigma^{\rm Born}$\,(pb) \\
            \hline
            &4.189  & 570.0  &0.668  &1.056  &$   32.0\pm  5.5$  & 1.1  &$   8.3\pm 1.4$  & 1.6  &$ 44.9\pm 7.8\pm 3.5$ \\
            &4.199  & 526.6  &0.677  &1.056  &$   62.6\pm  8.6$  & 2.0  &$  13.4\pm 1.8$  & 2.6  &$ 48.5\pm 6.7\pm 3.8$ \\
            &4.209  & 572.1  &0.697  &1.057  &$  193.3\pm 14.2$  & 2.9  &$  37.5\pm 2.7$  & 3.3  &$ 94.4\pm 6.9\pm 7.5$ \\
            &4.219  & 569.2  &0.725  &1.056  &$  291.0\pm 17.1$  & 3.6  &$  55.0\pm 3.2$  & 4.0  &$110.2\pm 6.5\pm 8.7$ \\      
            &4.226  &1111.9  &0.749  &1.056  &$ 1173.5\pm 42.1$  & 4.3  &$ 173.9\pm 6.2$  & 4.8  &$145.6\pm 5.2\pm11.5$ \\   
            &4.236  & 530.3  &0.770  &1.056  &$  468.0\pm 24.1$  & 4.2  &$  88.5\pm 4.6$  & 4.8  &$151.7\pm 7.8\pm11.5$ \\
            &4.242  &  55.9  &0.780  &1.055  &$   51.6\pm  8.9$  & 5.4  &$   9.7\pm 1.7$  & 6.1  &$122.3\pm21.2\pm 8.5$ \\     
            &4.244  & 538.1  &0.784  &1.056  &$  522.4\pm 26.6$  & 5.3  &$  98.5\pm 5.0$  & 6.0  &$130.4\pm 6.6\pm 9.0$ \\ 
            &4.258  & 828.4  &0.795  &1.054  &$ 1077.9\pm 37.9$  & 5.5  &$ 162.8\pm 5.7$  & 6.2  &$132.6\pm 4.7\pm10.8$ \\
            &4.267  & 531.1  &0.799  &1.053  &$  571.3\pm 28.2$  & 4.6  &$ 109.5\pm 5.4$  & 5.3  &$163.6\pm 8.1\pm11.6$ \\
            &4.278  & 175.7  &0.803  &1.053  &$  177.7\pm 16.1$  & 4.6  &$  34.1\pm 3.1$  & 5.2  &$153.7\pm13.9\pm10.9$ \\
            &4.287  & 494.2  &0.803  &1.053  &$  578.5\pm 27.7$  & 5.8  &$ 109.9\pm 5.3$  & 6.5  &$141.8\pm 6.8\pm10.4$ \\
            &4.308  &  45.1  &0.802  &1.052  &$   85.0\pm 11.1$  & 5.5  &$  16.3\pm 2.1$  & 6.3  &$239.0\pm31.3\pm19.1$ \\    
            &4.311  & 494.3  &0.802  &1.052  &$  745.5\pm 33.2$  & 5.5  &$ 143.8\pm 6.4$  & 6.3  &$193.7\pm 8.6\pm15.5$ \\
            &4.337  & 506.1  &0.798  &1.051  &$ 1078.9\pm 42.2$  & 7.4  &$ 201.5\pm 7.9$  & 8.2  &$203.7\pm 8.0\pm19.6$ \\
            &4.358  & 543.9  &0.798  &1.051  &$ 1604.8\pm 50.2$  & 8.0  &$ 304.4\pm 9.5$  & 9.0  &$262.3\pm 8.2\pm20.1$ \\    
            &4.377  & 524.7  &0.795  &1.051  &$ 1909.7\pm 55.2$  & 6.0  &$ 367.6\pm10.6$  & 6.8  &$431.5\pm12.5\pm35.9$ \\ 
            &4.387  &  55.6  &0.794  &1.051  &$  234.1\pm 19.2$  & 7.8  &$  45.2\pm 3.7$  & 8.9  &$384.3\pm31.5\pm28.0$ \\
            &4.395  & 508.2  &0.792  &1.051  &$ 2421.6\pm 60.0$  & 7.6  &$ 466.1\pm11.5$  & 8.7  &$446.5\pm11.1\pm32.5$ \\   
            &4.416  &1090.7  &0.794  &1.052  &$ 6225.0\pm 95.7$  & 7.4  &$1185.2\pm18.2$  & 8.3  &$547.7\pm 8.4\pm38.5$ \\
            &4.436  & 570.6  &0.796  &1.054  &$ 3798.9\pm 74.6$  & 7.6  &$ 755.4\pm14.8$  & 9.0  &$619.3\pm12.2\pm42.2$ \\
            &4.467  & 111.1  &0.810  &1.055  &$  929.9\pm 35.9$  & 7.8  &$ 176.2\pm 6.8$  & 8.8  &$742.9\pm28.7\pm50.8$ \\
            &4.527  & 112.1  &0.863  &1.054  &$ 1128.2\pm 39.2$  & 8.4  &$ 217.5\pm 7.5$  & 9.7  &$776.7\pm27.0\pm60.5$ \\
            &4.575  &  48.9  &0.900  &1.054  &$  430.3\pm 25.1$  &10.2  &$  86.1\pm 5.0$  &12.1  &$537.1\pm31.4\pm36.4$ \\
            &4.600  & 586.9  &0.905  &1.055  &$ 5964.6\pm 95.1$  &10.4  &$1206.9\pm19.2$  &12.5  &$606.4\pm 9.7\pm41.2$ \\   
            &4.613  & 103.7  &0.908  &1.055  &$ 1019.2\pm 38.9$  & 9.9  &$ 210.1\pm 8.0$  &12.1  &$613.4\pm23.4\pm41.2$ \\
            &4.628  & 521.5  &0.903  &1.054  &$ 5115.1\pm 88.6$  &10.2  &$1026.4\pm17.8$  &12.1  &$599.2\pm10.4\pm40.5$ \\
            &4.641  & 551.7  &0.900  &1.054  &$ 5404.6\pm 90.2$  &10.0  &$1123.5\pm18.7$  &12.4  &$608.4\pm10.1\pm44.4$ \\   
            &4.661  & 529.4  &0.893  &1.054  &$ 5434.9\pm 91.8$  &10.1  &$1111.6\pm18.8$  &12.2  &$641.7\pm10.8\pm47.1$ \\
            &4.682  &1667.4  &0.894  &1.054  &$18143.7\pm168.0$  &10.6  &$3743.9\pm34.7$  &12.9  &$647.2\pm 6.0\pm45.3$ \\
            &4.699  & 535.5  &0.900  &1.055  &$ 6088.0\pm 97.2$  &10.4  &$1265.0\pm20.2$  &12.9  &$680.3\pm10.9\pm48.9$ \\    
            &4.740  & 163.9  &0.925  &1.055  &$ 1761.7\pm 54.3$  &11.0  &$ 374.8\pm11.5$  &13.9  &$595.0\pm18.3\pm43.3$ \\
            &4.750  & 366.6  &0.934  &1.055  &$ 3902.4\pm 80.9$  &11.0  &$ 824.4\pm17.1$  &13.8  &$580.0\pm12.0\pm40.9$ \\
            &4.781  & 511.5  &0.957  &1.055  &$ 5185.2\pm 95.2$  &10.9  &$1113.7\pm20.4$  &13.9  &$546.8\pm10.0\pm36.6$ \\
            &4.843  & 525.2  &0.979  &1.056  &$ 4539.6\pm 91.7$  &10.6  &$ 972.9\pm19.7$  &13.4  &$468.4\pm 9.5\pm33.1$ \\
            &4.918  & 207.8  &0.973  &1.056  &$ 1674.8\pm 55.2$  &10.4  &$ 370.3\pm12.2$  &13.6  &$448.7\pm14.8\pm31.8$ \\
            &4.951  & 159.3  &0.965  &1.056  &$ 1273.9\pm 39.5$  &10.3  &$ 283.3\pm 8.8$  &13.5  &$452.9\pm14.1\pm31.1$ \\
            \hline\hline
            \end{tabular}
            \label{supp:tab:xs_xyz}
        \end{center}
    \end{table}
    
    \begin{table}[htp]
        \begin{center}
            \caption{
                The Born cross section of $\ee\to\dstzero\dstminus\pip$ for Scan data sets, where the charge conjugation mode is also included.
                The first uncertainties are statistical and the second ones systematic.
                }
            \begin{tabular}{c c c c c c c c c c}
            \hline\hline
            &$\sqrt{s}$\,(GeV) &$\mathcal{L}_{\rm int}$\,(pb$^{-1}$) &$1+\delta^{\rm ISR}$ &$\frac{1}{|1-\Pi|^{2}}$ &$N_{D^{0}\text{tag}}^{\rm obs}$ &$\epsilon_{D^{0}\text{tag}}$\,(\%) &$N_{D^{-}\text{tag}}^{\rm obs}$ &$\epsilon_{D^{-}\text{tag}}$\,(\%) &$\sigma^{\rm Born}$\,(pb) \\
            \hline
            &4.200  &6.8  &0.677  &1.057  &$  2.9\pm1.8 $  &2.3  &$0.6 \pm0.4$  & 2.8  &$156.2\pm 94.9 \pm12.3$ \\
            &4.203  &7.6  &0.681  &1.057  &$  3.4\pm1.9 $  &2.5  &$0.7 \pm0.4$  & 3.0  &$146.3\pm 82.4 \pm11.6$ \\
            &4.207  &7.7  &0.691  &1.057  &$  3.7\pm2.1 $  &2.9  &$0.7 \pm0.4$  & 3.3  &$134.3\pm 78.2 \pm10.6$ \\
            &4.212  &7.8  &0.705  &1.057  &$  4.2\pm2.2 $  &3.2  &$0.8 \pm0.4$  & 3.7  &$134.6\pm 69.6 \pm10.6$ \\
            &4.217  &7.9  &0.720  &1.056  &$  6.0\pm2.5 $  &3.8  &$1.1 \pm0.5$  & 4.2  &$158.1\pm 65.7 \pm12.5$ \\
            &4.222  &8.2  &0.738  &1.056  &$  6.5\pm3.1 $  &4.0  &$1.2 \pm0.6$  & 4.4  &$150.2\pm 70.8 \pm11.9$ \\
            &4.227  &8.2  &0.751  &1.056  &$  5.9\pm2.6 $  &4.2  &$1.1 \pm0.5$  & 4.7  &$128.8\pm 56.6 \pm10.2$ \\
            &4.232  &8.3  &0.762  &1.056  &$  7.0\pm3.0 $  &4.2  &$1.3 \pm0.6$  & 4.7  &$148.8\pm 64.0 \pm11.8$ \\
            &4.237  &7.8  &0.773  &1.055  &$  3.5\pm2.9 $  &5.0  &$0.7 \pm0.5$  & 5.6  &$ 65.0\pm 53.7 \pm4.5 $ \\
            &4.240  &8.6  &0.778  &1.055  &$ 10.0\pm4.1 $  &5.2  &$1.8 \pm0.8$  & 5.6  &$163.5\pm 67.2 \pm11.3$ \\
            &4.242  &8.5  &0.781  &1.056  &$  8.1\pm3.3 $  &5.5  &$1.5 \pm0.6$  & 5.8  &$125.4\pm 51.2 \pm8.7 $ \\
            &4.245  &8.6  &0.784  &1.056  &$ 11.7\pm3.7 $  &5.5  &$2.2 \pm0.7$  & 6.0  &$179.1\pm 57.1 \pm12.4$ \\
            &4.247  &8.6  &0.787  &1.055  &$ 14.9\pm3.7 $  &5.6  &$2.7 \pm0.7$  & 6.1  &$219.5\pm 55.3 \pm15.2$ \\
            &4.252  &8.7  &0.792  &1.054  &$  5.3\pm2.9 $  &5.3  &$1.0 \pm0.5$  & 5.9  &$ 81.9\pm 45.1 \pm6.7 $ \\
            &4.257  &8.9  &0.795  &1.054  &$ 10.7\pm4.3 $  &5.4  &$2.1 \pm0.8$  & 6.1  &$157.6\pm 62.8 \pm12.8$ \\
            &4.262  &8.6  &0.796  &1.053  &$ 16.9\pm3.9 $  &4.6  &$3.2 \pm0.7$  & 5.1  &$300.1\pm 68.9 \pm21.2$ \\
            &4.267  &8.6  &0.798  &1.053  &$  5.2\pm3.4 $  &4.7  &$1.0 \pm0.6$  & 5.2  &$ 90.8\pm 60.1 \pm6.4 $ \\
            &4.272  &8.6  &0.800  &1.053  &$  7.7\pm3.2 $  &4.7  &$1.5 \pm0.6$  & 5.4  &$133.5\pm 55.9 \pm9.4 $ \\
            &4.277  &8.7  &0.800  &1.053  &$ 10.2\pm4.3 $  &5.9  &$1.9 \pm0.8$  & 6.7  &$139.0\pm 57.7 \pm9.8 $ \\
            &4.282  &8.6  &0.803  &1.053  &$ 16.0\pm3.9 $  &6.1  &$3.0 \pm0.7$  & 6.7  &$214.0\pm 52.4 \pm15.7$ \\
            &4.287  &9.0  &0.802  &1.053  &$  9.1\pm4.0 $  &6.1  &$1.7 \pm0.8$  & 6.8  &$117.6\pm 51.4 \pm8.6 $ \\
            &4.297  &8.5  &0.801  &1.052  &$ 15.9\pm4.9 $  &5.5  &$2.9 \pm0.9$  & 6.0  &$242.0\pm 74.9 \pm19.3$ \\
            &4.307  &8.6  &0.803  &1.052  &$ 10.9\pm4.2 $  &5.4  &$2.1 \pm0.8$  & 6.2  &$163.2\pm 62.5 \pm13.0$ \\
            &4.317  &9.3  &0.801  &1.052  &$  3.7\pm3.6 $  &5.9  &$0.8 \pm0.7$  & 6.5  &$ 53.4\pm 46.1 \pm4.3 $ \\
            &4.327  &8.7  &0.800  &1.051  &$ 30.2\pm6.9 $  &7.4  &$5.7 \pm1.3$  & 8.2  &$333.1\pm 75.9 \pm32.0$ \\
            &4.337  &8.7  &0.798  &1.051  &$ 18.2\pm6.0 $  &7.4  &$3.4 \pm1.1$  & 8.2  &$199.9\pm 65.6 \pm19.2$ \\
            &4.347  &8.5  &0.798  &1.051  &$ 23.8\pm6.4 $  &7.8  &$4.3 \pm1.2$  & 8.4  &$251.9\pm 68.0 \pm19.3$ \\
            &4.357  &8.1  &0.797  &1.051  &$ 36.9\pm7.2 $  &7.6  &$6.9 \pm1.3$  & 8.4  &$428.1\pm 83.0 \pm32.8$ \\
            &4.367  &8.5  &0.796  &1.051  &$ 23.0\pm7.2 $  &6.1  &$4.2 \pm1.3$  & 6.6  &$316.5\pm 99.5 \pm26.3$ \\
            &4.377  &8.2  &0.796  &1.051  &$ 29.9\pm7.1 $  &6.2  &$5.7 \pm1.4$  & 7.0  &$421.9\pm 99.7 \pm35.1$ \\
            &4.387  &7.5  &0.794  &1.051  &$ 40.2\pm8.1 $  &7.9  &$7.3 \pm1.5$  & 8.5  &$487.6\pm 98.4 \pm35.5$ \\
            &4.392  &7.4  &0.794  &1.051  &$ 37.5\pm7.7 $  &7.9  &$7.2 \pm1.5$  & 8.9  &$454.6\pm 93.3 \pm33.1$ \\
            &4.397  &7.2  &0.793  &1.051  &$ 38.3\pm7.8 $  &8.0  &$7.2 \pm1.5$  & 9.0  &$471.3\pm 96.1 \pm34.3$ \\
            &4.407  &6.4  &0.793  &1.052  &$ 36.3\pm7.7 $  &7.3  &$6.9 \pm1.5$  & 8.2  &$558.4\pm 118.7\pm39.2$ \\
            &4.417  &7.5  &0.793  &1.052  &$ 37.4\pm8.2 $  &7.5  &$7.1 \pm1.6$  & 8.4  &$473.1\pm 103.7\pm33.2$ \\
            &4.422  &7.4  &0.793  &1.052  &$ 43.9\pm8.2 $  &7.4  &$8.3 \pm1.6$  & 8.3  &$564.9\pm 106.1\pm39.7$ \\
            &4.427  &6.8  &0.794  &1.053  &$ 42.0\pm8.0 $  &8.7  &$8.1 \pm1.5$  & 9.9  &$503.3\pm 95.9 \pm34.3$ \\
            &4.437  &7.6  &0.796  &1.054  &$ 52.5\pm9.3 $  &8.0  &$10.3\pm1.8$  & 9.3  &$609.9\pm 108.1\pm41.6$ \\
            &4.447  &7.7  &0.800  &1.054  &$ 67.7\pm9.8 $  &7.3  &$12.8\pm1.9$  & 8.2  &$845.9\pm 123.1\pm57.7$ \\
            &4.457  &8.7  &0.803  &1.055  &$ 69.9\pm10.3$  &7.4  &$13.5\pm2.0$  & 8.5  &$756.7\pm 111.1\pm51.7$ \\
            &4.477  &8.2  &0.816  &1.055  &$ 85.4\pm11.0$  &7.9  &$15.6\pm2.0$  & 8.6  &$909.9\pm 117.0\pm62.2$ \\
            &4.497  &8.0  &0.833  &1.055  &$ 81.6\pm10.8$  &7.9  &$15.6\pm2.1$  & 8.9  &$874.1\pm 116.0\pm59.7$ \\
            &4.517  &8.7  &0.855  &1.055  &$ 73.6\pm10.8$  &8.2  &$13.9\pm2.0$  & 9.1  &$682.5\pm 100.3\pm53.2$ \\
            &4.537  &9.3  &0.875  &1.054  &$101.3\pm12.3$  &9.4  &$20.1\pm2.4$  &11.1  &$739.8\pm 89.7 \pm57.6$ \\
            &4.547  &8.8  &0.884  &1.054  &$ 92.0\pm11.8$  &9.6  &$18.2\pm2.3$  &11.2  &$697.0\pm 89.7 \pm54.3$ \\
            &4.557  &8.3  &0.892  &1.054  &$ 90.9\pm11.3$  &9.8  &$17.9\pm2.2$  &11.5  &$704.7\pm 87.8 \pm47.8$ \\
            &4.567  &8.4  &0.897  &1.054  &$ 96.5\pm12.3$  &9.8  &$19.0\pm2.4$  &11.4  &$737.6\pm 94.2 \pm50.0$ \\
            &4.577  &8.5  &0.901  &1.055  &$ 78.7\pm11.3$  &9.7  &$15.8\pm2.3$  &11.6  &$591.7\pm 84.7 \pm40.1$ \\
            &4.587  &8.2  &0.905  &1.055  &$ 83.3\pm11.6$  &9.9  &$16.6\pm2.3$  &11.7  &$639.5\pm 89.1 \pm43.4$ \\
            \hline\hline
            \end{tabular}
            \label{supp:tab:xs_scan}
        \end{center}
    \end{table}

    \clearpage
    \section{Multiple solutions of lineshape fit}
    
    The dressed cross section is parameterized as the coherent sum of a continuum amplitude for $\ee\to\dstzero\dstminus\pip$ and three $R_{k}\to\dstzero\dstminus\pip$ resonance amplitudes:
    \begin{eqnarray}
    \sigma^{\rm dressed}(\sqrt{s})=C_{0}|C_{1}\sqrt{\Phi(\sqrt{s})}+\sum^{3}_{k=1}\text{BW}_{k}(\sqrt{s})e^{i\phi_{k}}|^2,\nonumber
    \end{eqnarray}
    where the relativistic Breit-Wigner functions are given by
    \begin{eqnarray}
    \text{BW}_{k}(\sqrt{s})=\frac{m_{k}}{\sqrt{s}}\frac{\sqrt{12\pi\Gamma^{ee}_{k}\mathcal{B}_{k}\Gamma^{\rm tot}_{k}}}{s-m_{k}^{2}+im_{k}\Gamma^{\rm tot}_{k}}\sqrt{\frac{\Phi(\sqrt{s})}{\Phi(m_{k})}},\nonumber
    \end{eqnarray}
    and the 3-body phase space contribution $\Phi(\sqrt{s})=\iint\frac{1}{(2\pi)^{3}32(\sqrt{s})^3}dm_{23}^{2}dm_{12}^{2}$.
    All the parameters in the fit are free except for $C_{0}=3.894\times10^{5}\,\unit{nb}\gev^{2}$ as a unit conversion factor.
    
\begin{figure}[htp]
        \begin{center}
            \includegraphics[width=0.49\linewidth]{./plot/Solution_0.eps}\put(-50,75){(I)}
            \includegraphics[width=0.49\linewidth]{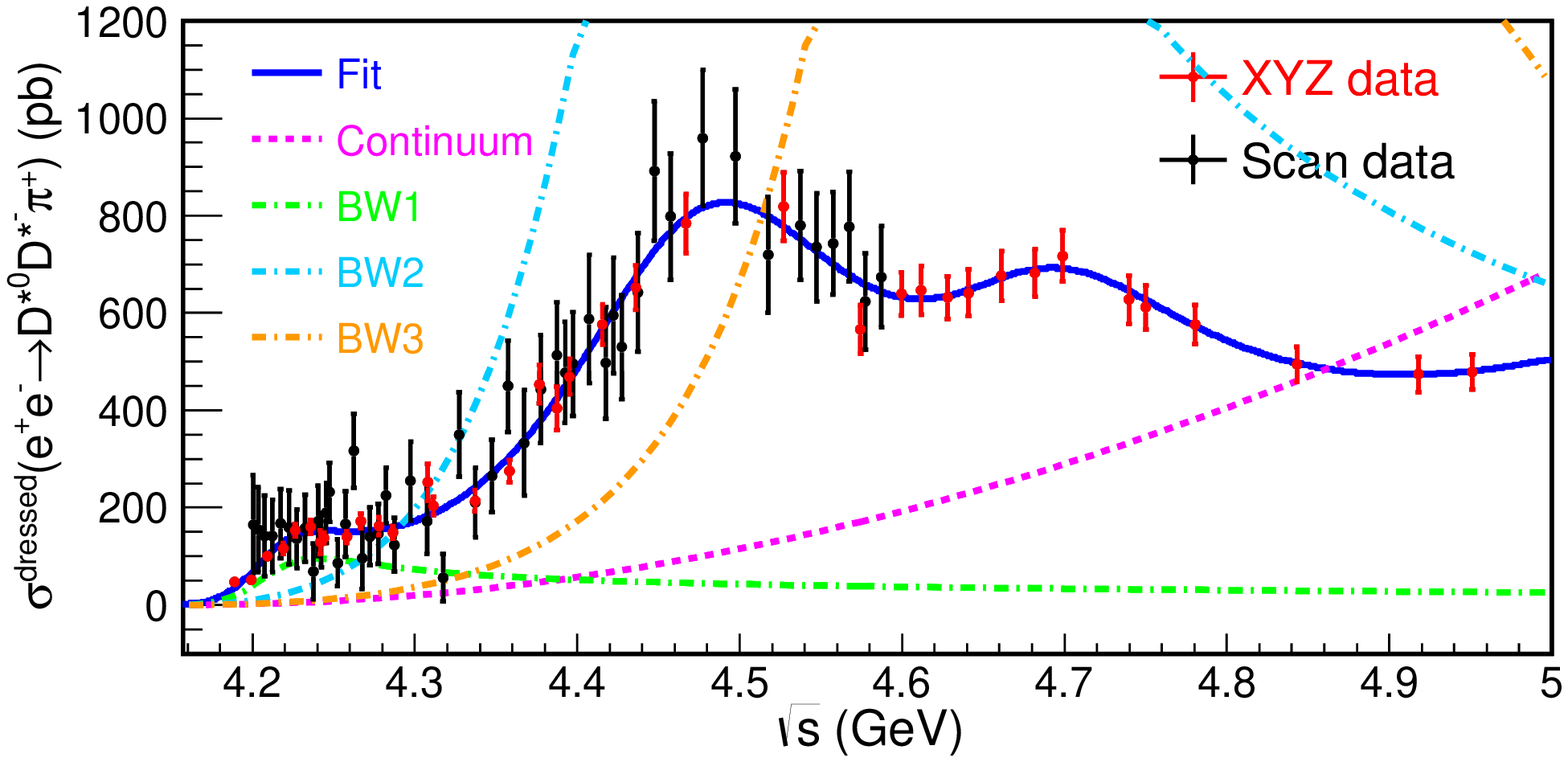}\put(-50,75){(II)}\\
            \includegraphics[width=0.49\linewidth]{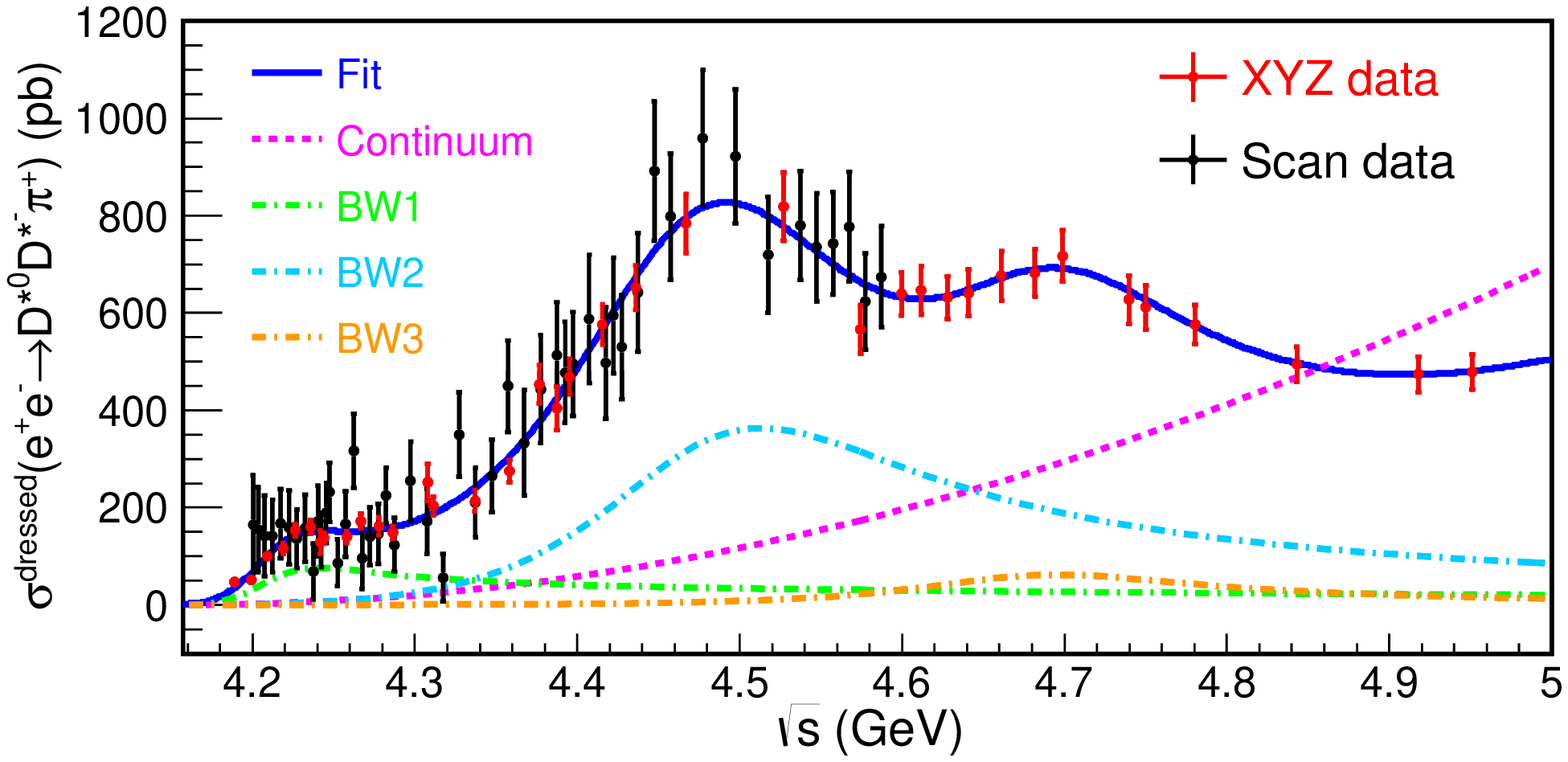}\put(-50,75){(III)}
            \includegraphics[width=0.49\linewidth]{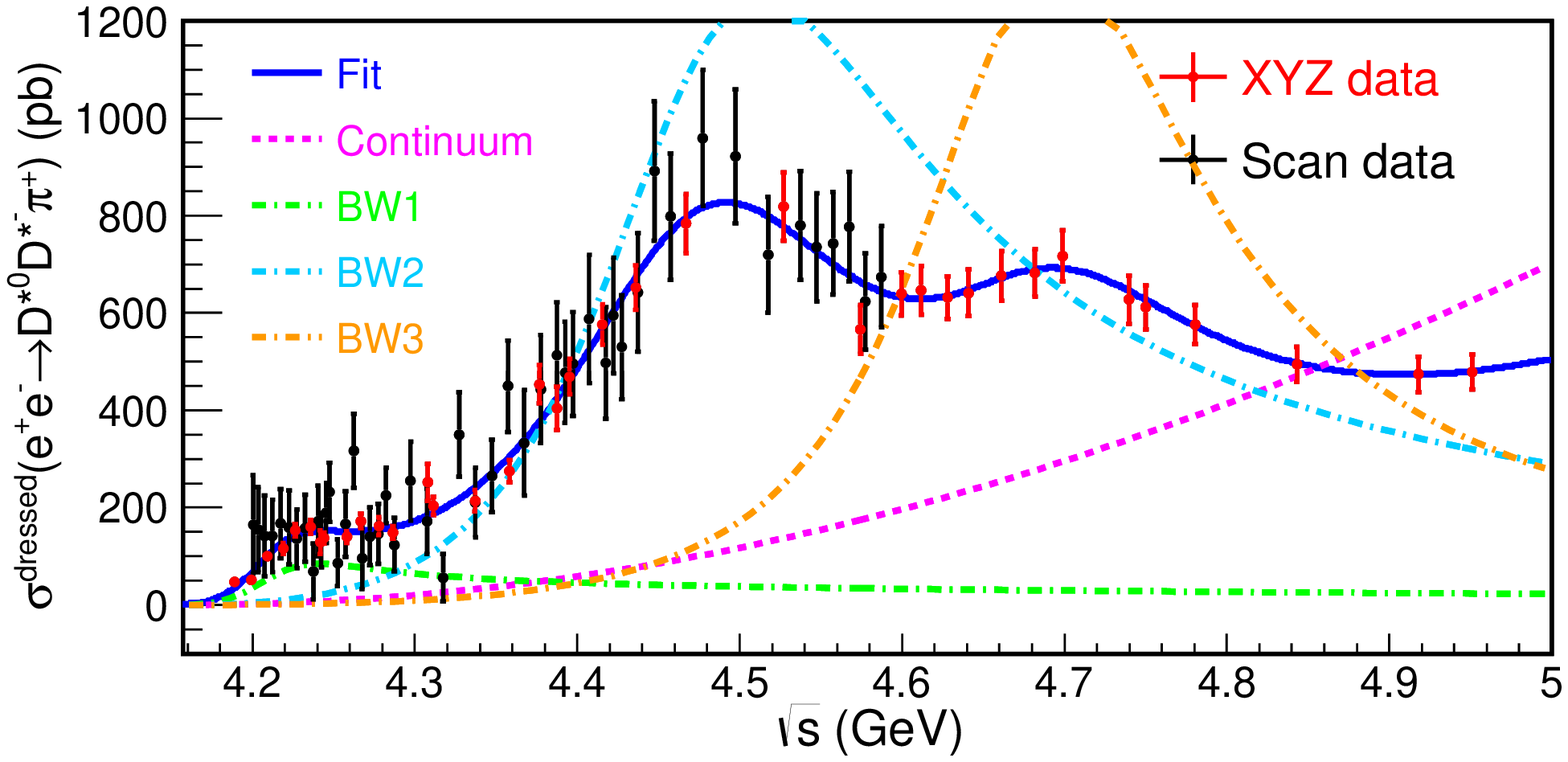}\put(-50,75){(IV)}\\
            \includegraphics[width=0.49\linewidth]{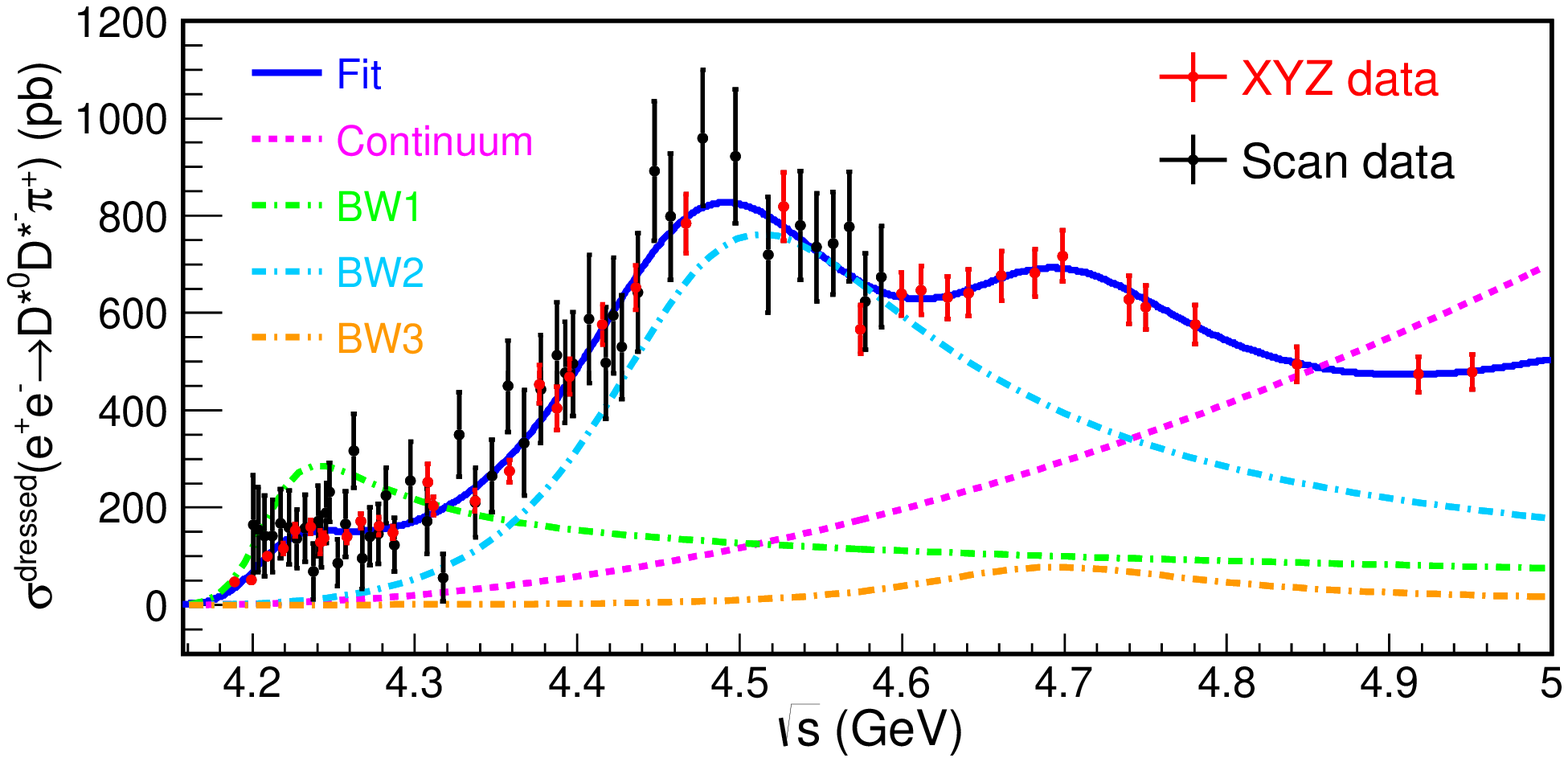}\put(-50,75){(V)}
            \includegraphics[width=0.49\linewidth]{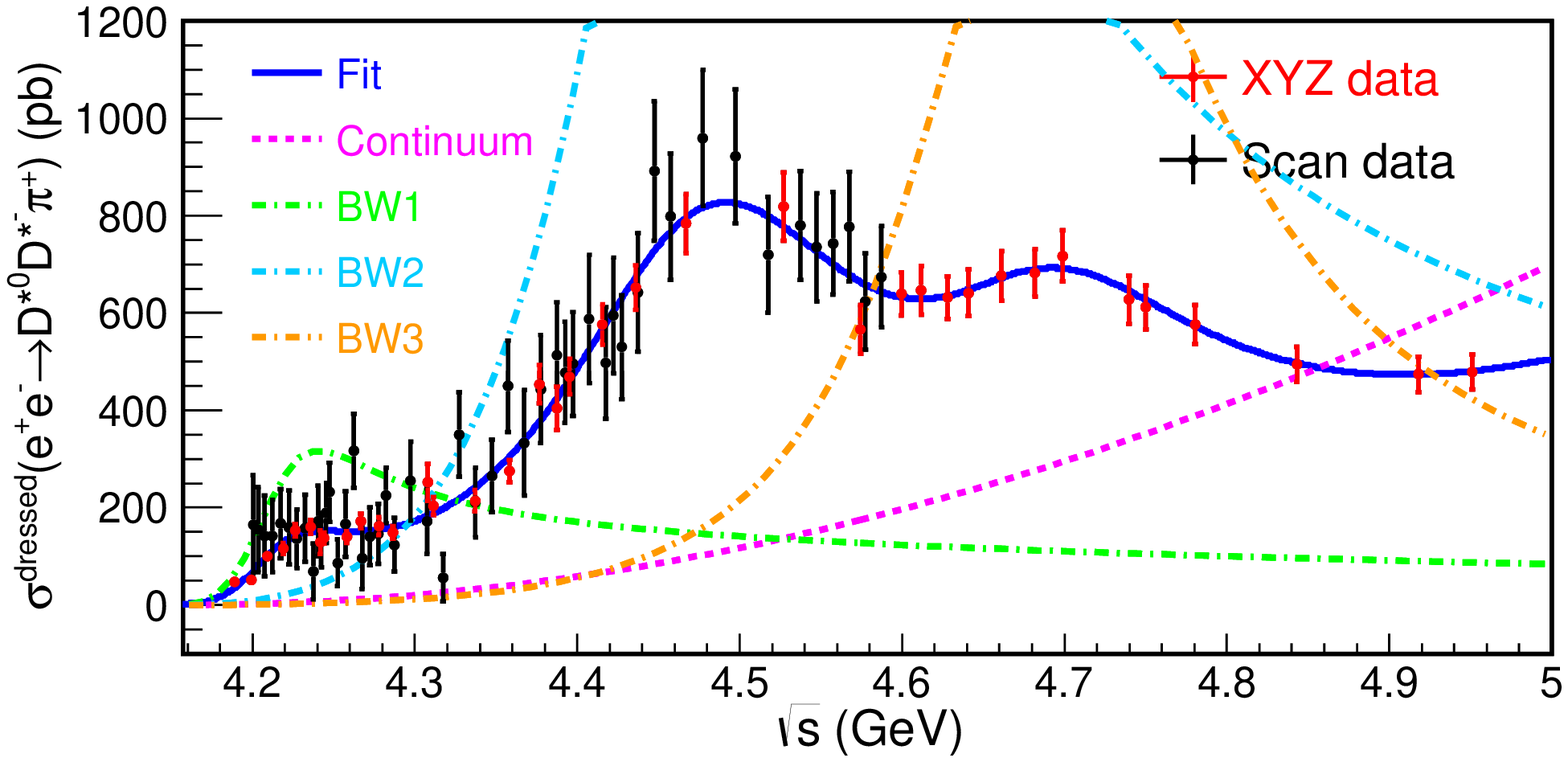}\put(-50,75){(VI)}\\
             \includegraphics[width=0.49\linewidth]{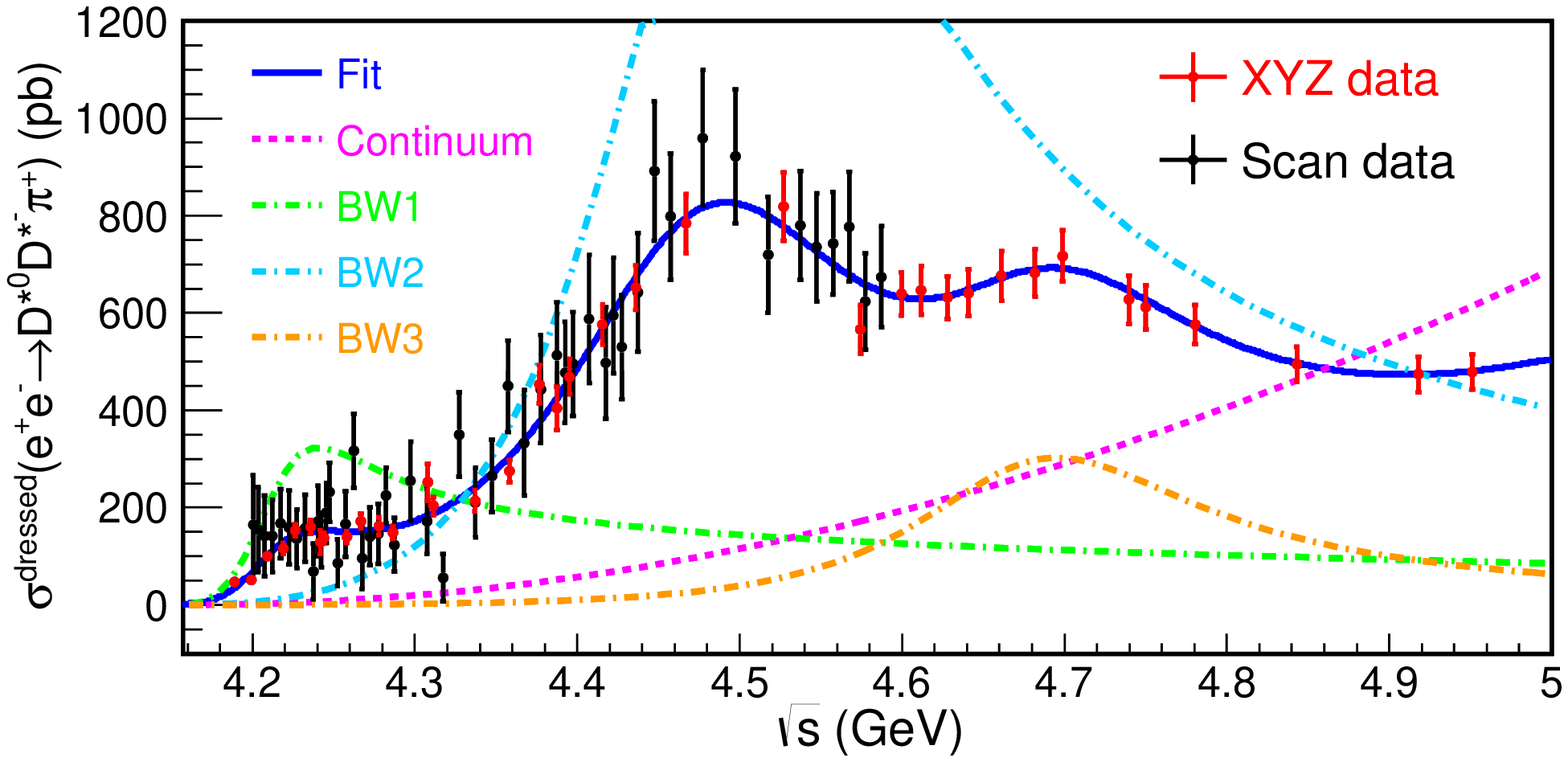}\put(-50,75){(VII)}
             \includegraphics[width=0.49\linewidth]{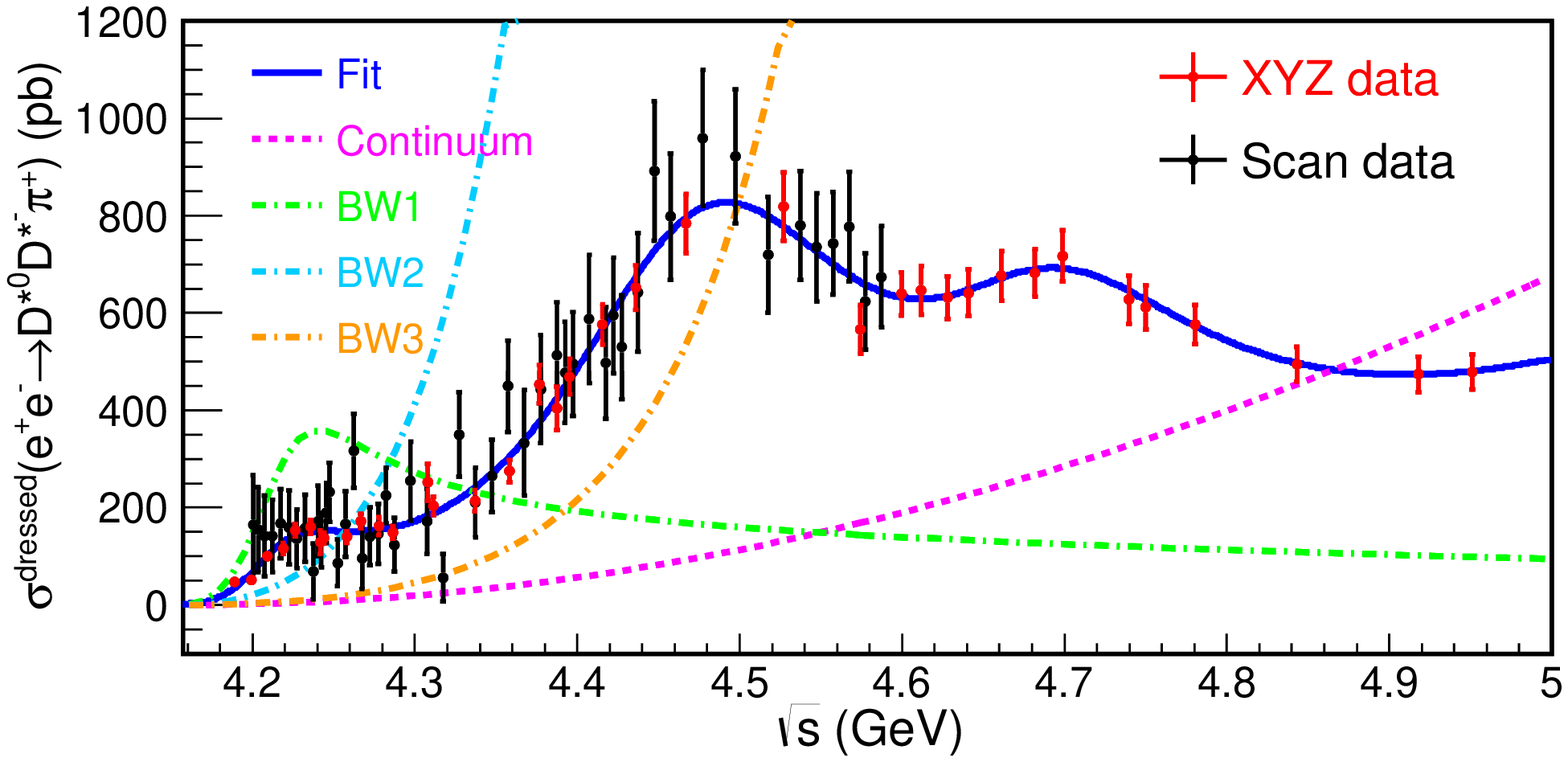}\put(-50,75){(VIII)}
            \caption{The fit results of the dressed cross section lineshape of $\ee\to\dstzero\dstminus\pip$, where the charge conjugation mode is also included. The black and red points with error bars are data, including statistical and systematic uncertainties. The blue curve is the total fit. The green, azure and orange dashed curves describe three BW functions, and the pink dashed-curve is the three body phase space contribution.}
            \label{supp:fig:fit_LS}
        \end{center}
    \end{figure}

    In the above function, mathematically there exists eight solutions with the same outputs of the dressed cross sections~\cite{Bai:2019jrb}, which is due to the interference between any two of the involved components in the fitting function.
    The different combinations of the amplitude and relative phase angle of each component can lead to the same outcome, whose sizes are presented by the parameters $\Gamma^{ee}_{k}\mathcal{B}_{k}$ and $\phi_k$, respectively.
    This means that there are eight degenerate solutions with the same fit quality, as shown in Fig.~\ref{supp:fig:fit_LS}.
    Among them, the results of the resonance masses $m_k$ and total widths $\Gamma^{\rm tot}_{k}$ in the multiple solutions are equal, while those of the partial width and relative phase angle vary.

    
    \clearpage
    \section{Systematic uncertainties}
    The systematic uncertainty studies are performed at energy points where the signal yield is larger than 500 events.
    For other points suffering limited statistics, the uncertainty from the closest point is taken as its systematic uncertainty. 
    All the systematic uncertainties are studied on each tag method separately and then combined together according to their yields.
    To consider all the related systematic uncertainties in the measurement of the Born cross sections, the sources of systematic uncertainty are divided into three categories.
    
    The first category includes uncertainties associated with the detection efficiencies, such as
    tracking, particle identification, $\pizero$ reconstruction,
    signal region requirements of the reconstructed unstable particles ({\it i.e.}, the $P^{*}(\pizero)$ momentum rejection region, $\dzero(\dminus)$ and $\dstzero(\dstminus)$ mass window requirements), MC simulation model and ISR correction factors.
    The uncertainty of detection and PID efficiency is 1.0\% for each charged track~\cite{BESIII:2011ysp}, and 2.0\% for $\pizero$ reconstruction~\cite{BESIII:2010ank}.
    The uncertainties associated with the $P^{*}(\pizero)$, $M(\dzero(\dminus))$ and $M(\dstzero(\dstminus))$ windows are estimated by re-extracting the detection efficiencies with Gaussian-smeared MC samples where the Gaussian parameters are obtained from the discrepancies between data and MC simulation.
    The MC samples are corrected according to the partial-wave-analysis (PWA) results at each energy point.
    To estimate the uncertainties from PWA-corrected MC samples, the samples with different PWA results are re-corrected by changing the possible components used in the PWA.
    The differences of efficiencies extracted from nominal and re-corrected MC samples are taken as the systematic uncertainty of the signal model.
    The ISR correction factors can have an impact on the detection efficiencies by affecting the slope of the lineshape, and hence, they are treated together as $(1+\delta^{\rm ISR})\epsilon$.
    Since the ISR correction factors are estimated by the fitting-iteration method, the uncertainty of this item comes from four different parts:
    the differences between the last two iterations are taken as the uncertainty of the iteration method itself;
    the uncertainty from the lineshape fitting model used in the iteration are estimated by replacing the phase space model with a parameterized function;
    the uncertainty of $(1+\delta^{\rm ISR})\epsilon$ is estimated by 500 groups of cross-section toys which are re-sampled according to the uncertainty of the parameters and the corresponding covariance matrix;
    the vacuum polarization factors are taken from QED calculations at each energy point and affect the slope of the lineshape with an estimated uncertainty of $0.5\%$.
    
    The second category includes uncertainties associated with the signal shape, background shape and fit range, which affect the estimation of signal yields.
    The uncertainty of the signal shape is estimated by convolving it with a double-Gaussian function in the fit instead of a single Gaussian function in the nominal results.
    The description of the background shape is changed from a $2^{\rm nd}$ order Chebyshev function to a linear function in the fit and the differences between the two cases are considered as the uncertainty.
    The uncertainty associated with the fit range uncertainty is determined by altering the fit range from $(1.91, 2.12)\gevcc$ to $(1.911, 2.12)\gevcc$. 
    
    The last category includes the luminosity and quoted BFs, where the former uncertainty is $1.0\%$ at each energy point~\cite{ref:lumi_eng_scan:BESIII:2017lkp,ref:lumi_eng_xyz_1:BESIII:2020eyu,ref:lumi_eng_xyz_2:BESIII:2022ulv}, and the latter uncertainties are taken from the PDG~\cite{Workman:2022ynf}.
    
    Assuming no significant correlations between sources, the total systematic uncertainty is obtained as the sum in quadrature.
    Table~\ref{supp:tab:total_sys} summarizes the systematic uncertainties of the cross section at various energy points. 
    
        \begin{table}[h]
            \begin{center}
            \caption{
                The summary of all the systematic uncertainties (\%) in the Born cross section measurement.
                The items marked with `$\dagger$' are treated as fully correlated uncertainties, while others are uncorrelated uncertainties.
                }
                \begin{tabular}{c|c c c c c c c c c c c|c}
                    \hline\hline
                    $\sqrt{s}$\,(GeV)   &Track$^{\dagger}$   &PID$^{\dagger}$   &$\pizero$$^{\dagger}$   
                        &Signal region       &Decay Model       &$(1+\delta^{\rm ISR})^{\dagger}$
                        &Signal shape        &Bkg. shape        &Fit range  
                        &$\mathcal{L}_{\rm int}^{\dagger}$      &$\mathcal{B}^{\dagger}$
                        &Total\\\hline
                    4.226   &3.7   &3.7   &2.8   &0.5   &0.9   &1.4   &2.5   &0.9   &3.1   &1.0   &2.7   &7.9\\
                    4.236   &3.7   &3.7   &2.7   &0.4   &0.3   &1.0   &2.3   &0.5   &2.6   &1.0   &2.7   &7.6\\
                    4.244   &3.6   &3.6   &2.8   &0.6   &0.6   &0.8   &1.0   &1.2   &1.3   &1.0   &2.7   &6.9\\
                    4.258   &3.6   &3.6   &2.8   &0.6   &0.0   &0.7   &2.6   &2.5   &3.2   &1.0   &2.7   &8.1\\
                    4.267   &3.6   &3.6   &2.8   &0.7   &0.1   &0.7   &1.8   &0.5   &1.8   &1.0   &2.7   &7.1\\
                    4.288   &3.7   &3.7   &2.8   &0.2   &0.1   &1.9   &1.6   &1.6   &1.6   &1.0   &2.7   &7.3\\
                    4.312   &3.6   &3.6   &2.7   &0.3   &0.6   &2.9   &1.6   &3.0   &1.5   &1.0   &2.6   &8.0\\
                    4.337   &3.6   &3.6   &2.7   &0.4   &0.5   &2.3   &1.9   &6.0   &2.1   &1.0   &2.7   &9.6\\
                    4.358   &3.7   &3.7   &2.8   &0.5   &0.1   &2.3   &1.3   &2.5   &1.7   &1.0   &2.7   &7.7\\
                    4.377   &3.7   &3.7   &2.7   &0.4   &0.4   &1.2   &0.9   &4.6   &1.7   &1.0   &2.7   &8.3\\
                    4.397   &3.7   &3.7   &2.8   &0.3   &0.6   &0.6   &0.9   &2.7   &0.8   &1.0   &2.7   &7.3\\
                    4.416   &3.7   &3.7   &2.7   &0.6   &0.4   &0.8   &0.2   &2.3   &0.5   &1.0   &2.7   &7.0\\
                    4.436   &3.7   &3.7   &2.8   &0.3   &0.2   &0.9   &1.2   &1.3   &0.0   &1.0   &2.7   &6.8\\
                    4.467   &3.7   &3.7   &2.7   &0.5   &0.6   &1.0   &0.3   &1.5   &0.4   &1.0   &2.7   &6.8\\
                    4.527   &3.7   &3.7   &2.7   &0.4   &0.4   &0.9   &3.1   &1.0   &2.5   &1.0   &2.7   &7.8\\
                    4.575   &3.7   &3.7   &2.8   &0.5   &0.3   &0.6   &1.1   &0.6   &0.4   &1.0   &2.7   &6.8\\
                    4.600   &3.7   &3.7   &2.7   &0.5   &1.2   &0.9   &0.5   &0.3   &0.7   &1.0   &2.6   &6.8\\
                    4.612   &3.7   &3.7   &2.8   &0.2   &0.4   &1.1   &0.5   &0.7   &0.5   &1.0   &2.7   &6.7\\
                    4.628   &3.7   &3.7   &2.7   &0.4   &0.1   &1.3   &0.7   &0.9   &0.4   &1.0   &2.6   &6.8\\
                    4.641   &3.7   &3.7   &2.7   &0.3   &1.0   &2.7   &0.7   &1.1   &0.8   &1.0   &2.7   &7.3\\
                    4.661   &3.7   &3.7   &2.8   &0.3   &2.0   &2.6   &0.6   &0.3   &0.1   &1.0   &2.7   &7.3\\
                    4.681   &3.7   &3.7   &2.7   &0.3   &0.5   &2.4   &0.2   &0.5   &0.1   &1.0   &2.7   &7.0\\
                    4.698   &3.7   &3.7   &2.7   &0.2   &1.0   &2.5   &0.1   &1.1   &0.4   &1.0   &2.7   &7.2\\
                    4.740   &3.7   &3.7   &2.7   &0.5   &2.5   &1.5   &0.5   &0.8   &0.8   &1.0   &2.6   &7.3\\
                    4.750   &3.7   &3.7   &2.7   &0.3   &2.1   &1.0   &0.1   &0.6   &0.5   &1.0   &2.6   &7.0\\
                    4.781   &3.7   &3.7   &2.7   &0.4   &0.4   &0.8   &0.2   &0.4   &0.9   &1.0   &2.6   &6.7\\
                    4.843   &3.7   &3.7   &2.7   &0.2   &2.1   &1.2   &0.4   &0.9   &0.6   &1.0   &2.6   &7.1\\
                    4.918   &3.7   &3.7   &2.7   &0.4   &1.8   &1.2   &0.4   &1.4   &0.5   &1.0   &2.6   &7.1\\
                    4.951   &3.7   &3.7   &2.8   &0.4   &0.6   &0.8   &0.6   &1.6   &0.6   &1.0   &2.7   &6.9\\
                    \hline\hline
                \end{tabular}
            \label{supp:tab:total_sys}
        \end{center}
    \end{table}
\end{appendices}

\end{document}